\documentclass[preprint,aps,amsmath,nofootinbib,floatfix]{revtex4}
\usepackage{subfigure}
\usepackage{graphicx}
\usepackage{multirow}

\usepackage{color}
\usepackage{comment}

\usepackage{natbib}

\makeatletter

\newcommand{\Rmnum}[1]{\expandafter\@slowromancap\romannumeral #1@}
\makeatother
\newcommand{\met}{\ensuremath{{\not\mathrel{E}}_T}}\newcommand{\gev}{{\rm GeV}}

\def\sL{\ensuremath{\tilde t_L}}
\def\sR{\ensuremath{\tilde t_R}}
\def\bL{\ensuremath{\tilde b_L}}
\def\n{\ensuremath{\chi^0_1}}

\begin{document}

%\listoftables

%\date{\today}

\title{ {Complex decay chains of top and bottom squarks}}
\author{
         Jonathan Eckel$^{1}$\footnote{eckel@physics.arizona.edu},
        Shufang Su$^{1}$\footnote{shufang@email.arizona.edu},
         Huanian Zhang$^{1}$\footnote{fantasyzhn@email.arizona.edu}}

\affiliation{
$^{1}$ Department of Physics, University of Arizona, Tucson, Arizona  85721
}

\begin{abstract}

Current searches for the top squark mostly focus on the decay channels of $\tilde{t}_1 \rightarrow t \chi_1^0$ or $\tilde{t}_1 \rightarrow b \chi_1^\pm \rightarrow bW \chi_1^0$, leading to $tt/bbWW+\met$ final states for top squark pair production at the LHC.   In supersymmetric scenarios with light gauginos other than the neutralino lightest supersymmetric particle (LSP), different decay modes of the top squark could be dominant, which significantly weaken the current top squark search limits at the LHC.  Additionally, new decay modes offer alternative discovery channels for top squark searches.  In this paper, we study the top squark  and bottom squark decay in the Bino-like LSP case with light Wino or Higgsino next-to-LSPs (NLSPs), and identify cases in which additional decay modes become dominant.  We also perform a   collider analysis for top squark pair production with  mixed top squark decay final states of  $\tilde{t}_1 \to t {\chi}_2^0  \to th {\chi}_1^0$,  $\tilde{t}_1 \to b {\chi}_1^\pm  \to bW {\chi}_1^0 $, leading to the $bbbbjj\ell+\met$ collider signature.  The branching fraction for such decay varies between 25\% and 50\% for a top squark mass larger than 500 GeV with $M_2=M_1+150$ GeV.   At the 14 TeV LHC with 300 ${\rm fb}^{-1}$ integrated luminosity, the top squark can be excluded up to about 1040 GeV at the 95\% C.L., or be discovered up to 940 GeV at 5$\sigma$ significance. 

\end{abstract}

\maketitle

\section{Introduction}

The discovery of a 125 \gev~Higgs at the Large Hadron Collider (LHC) \cite{Aad:2012tfa, Chatrchyan:2012ufa} motivates the consideration of new  physics beyond the Standard Model (SM).   In the SM,   the Higgs receives unstable quadratically divergent  radiative corrections to its mass  from the top quark loop.  An unnatural cancellation is needed to recover the light physical Higgs mass, which is the so called ``Hierarchy problem''~\cite{Weinberg:1975gm}.  Supersymmetry (SUSY) provides a    solution to the naturalness problem by introducing superpartners to the SM particles, with interactions following the SUSY relations.  The quadratic divergence from the superpartners cancels that of the SM particles, with the remnant contributions being only  logarithmically divergent.   Given the   large top Yukawa coupling, the top and top squark (referred to as stop) sectors of the Minimal Supersymmetric Standard Model (MSSM) provide the largest radiative corrections to the Higgs mass.   Stop masses can not be too heavy in order  to avoid excessive fine tuning of the Higgs mass.   A TeV scale stop typically leads to fine tuning of about 1\%~\cite{Papucci:2011wy}.  Given the tight connection between the stop and Higgs sectors, it is important to fully explore the discovery potential of the stop at the LHC.

Most of the current searches for the light stop focus on the decay $\tilde{t}_1 \rightarrow t \chi_1^0$ or $\tilde{t}_1 \rightarrow b {\chi}_1^\pm  \to bW {\chi}_1^0 $,  with   $\chi_1^0$ being the stable lightest supersymmetric particle (LSP) appearing as missing energy ($\met$) at colliders.  For stop pair production at the LHC, such processes lead to  $tt+\met$ or $bbWW+\met$ final states.   However, due to the large SM  backgrounds from $t \bar t$, searches for  the stop can be  very challenging.   The current limits from ATLAS and CMS experiments   exclude stops with masses up to about 645 GeV for a light neutralino LSP~\cite{Aad:2014bva,Aad:2014kra,Aad:2014qaa,CMS:2013nia,CMS-PAS-SUS-14-011,Chatrchyan:2013xna}.  For small mass spitting between the stop and the LSP, $\tilde{t}_1\rightarrow c \chi_1^0$ and $\tilde{t}_1\rightarrow b f f^\prime \chi_1^0$ has been studied~\cite{Aad:2014nra, CMS-PAS-SUS-13-009}, with  limits for stop masses around 240 to 270 GeV.   Searches for the second stop with $\tilde{t}_2\rightarrow \tilde{t}_1 Z/h$ have also been performed, which provide stop mass limits around 540 to 600 GeV~\cite{Aad:2014mha,Khachatryan:2014doa,CMS-PAS-SUS-13-021}.   Stop decay with a gravitino LSP has also been studied in Ref.~\cite{Chatrchyan:2013mya,Aad:2014mha}.

Similarly, the current bottom squark (referred to as sbottom) searches mostly focus on $\tilde{b}_1 \rightarrow b \chi_1^0$, with $bb+\met$ being the dominant search channel.     Given data collected at the LHC 7/8 TeV, sbottoms with masses up to 700 GeV are excluded~\cite{Aad:2013ija,CMS-PAS-SUS-13-018,Aad:2014nra}.     Searches based on sbottom decay of    $\tilde{b}_1\rightarrow b \chi_2^0\rightarrow bZ/h\chi_1^0$ exclude sbottom masses between 340 and 600 GeV~\cite{Aad:2014lra,CMS-PAS-SUS-13-008}.   $\tilde{b}_1\rightarrow t \chi_1^\pm \rightarrow tW\chi_1^0$ decay has also been studied in multi-lepton final states~\cite{Aad:2014pda,Chatrchyan:2013fea,CMS-PAS-SUS-13-008,Chatrchyan:2014aea}, which excludes sbottom masses around 440 $-$590 GeV.   The left-handed sbottom mass is related to the left-handed stop mass since they are controlled by the same soft SUSY breaking mass parameter.  In this paper, we also study the left-handed sbottom decay patterns, as well as its collider signatures.  

There are other theoretical studies in the literature on the stop searches at the LHC, mostly focusing on the light stop decaying to light generation quarks \cite{Grober:2014aha,Agrawal:2013kha,Muhlleitner:2011ww,Aebischer:2014lfa,Boehm:1999tr} with little missing energy, which  mimics the $WW$ signal at the LHC \cite{Delgado:2012eu,Rolbiecki:2013fia,Curtin:2014zua,Kim:2014eva,Czakon:2014fka} or multi-$b$ jets final states from a light stop~\cite{Berenstein:2012fc}.    For the sbottom,   in a parameter space with highly degenerate sbottom and LSP masses,    a strategy has been proposed to search for sbottom based on boosting bottoms through an energetic initial radiation jet~\cite{Alvarez:2012wf}.  

The current stop and sbottom search limits, however,  could be significantly weakened when other   decay modes open, which could occur in many regions of MSSM parameter space.   On the other hand, the opening of new channels offers alternative discovery potential for stops and sbottoms  at the LHC.  It is thus  important to analyze all possible stop and sbottom  decay patterns to     fully explore the discovery potential at the 14 TeV LHC.

Even under the usual assumption of a Bino-like LSP, the existence of other light neutralino states, for example, Wino-like or Higgsino-like next-to-LSPs (NLSPs) could lead to new decay channels for the stop. For instance, $\tilde{t}_1$ could decay to $t \chi_{2,3}^0$, with $\chi_{2,3}^0$ further decaying to $Z\chi_1^0, h \chi_1^0$.   Given the relatively large ${\rm SU}(2)_L$ coupling and   top Yukawa coupling, compared to the ${\rm U}(1)_Y$ coupling relevant for the Bino-like LSP, decays to  $t \chi_{2,3}^0$ could even be dominant despite the phase space suppression.   In this paper, we study the stop and sbottom decay branching fractions for the Wino- or Higgsino-like NLSP case, considering the  minimal mixing and the maximal mixing scenarios in the stop sector, and outline the main search channels for the stops and bottoms at the LHC.

Given the discovery of the SM-like Higgs boson at the LHC, we can now use   final states with a Higgs boson to search for new physics beyond the SM.   To demonstrate the 14 TeV LHC reach with those complex stop  decay channels, we performed a  sample collider analysis with a Higgs in the final state:   $pp \rightarrow \tilde{t}_1\tilde{t}_1^*$ with mixed stop decay final states of  $\tilde{t}_1 \to t {\chi}_2^0  \to th {\chi}_1^0$,  $\tilde{t}_1 \to b {\chi}_1^\pm  \to bW {\chi}_1^0 $, leading to the $bbbbjj\ell+\met$ collider signature  with the assumption of  the branching fraction of $h\rightarrow b \bar{b}$ being the SM value of 57.7\%.   The branching fraction for such decay could vary between 25\% and 50\%  for a stop mass larger than 500 GeV with $M_2=M_1+150$ GeV.  By designing selection cuts to identify the signal while suppressing   SM backgrounds, we obtained the 95\% C.L. exclusion limit as well as the 5$\sigma$ discovery reach in $m_{\tilde{t}_1}$ versus $m_{\chi_1^0}$ plane at the 14 TeV LHC with 300 ${\rm fb}^{-1}$ integrated luminosity.     Note that other Higgs decay channels: $h\rightarrow WW, ZZ, \gamma\gamma, \tau\tau$ are not considered in our current analyses, which could lead to interesting multi-lepton final states or extremely clean (although suppressed)  $\gamma\gamma$ signatures.   Final states with $\chi_{2}^0 \rightarrow Z \chi_1^0$ are left for future studies.

The rest of the paper is organized as the following.
In Sec.~\ref{sec:MSSM_stop}, we present the third generation squark sector in the MSSM and discuss its connection to the Higgs sector.  In Sec.~\ref{sec:prod_decay}, we discuss the  stop and sbottom decays for various scenarios,   as well as the collider signatures for stop/sbottom pair production.   In Sec.~\ref{sec:limit}, we summarize the current  and future  LHC stop and sbottom search results from both ATLAS and CMS. In Sec.~\ref{sec:analyses},  we investigate the 14 TeV  reach  of the stop   via final states with a Higgs.  In Sec.~\ref{sec:conclusion}, we conclude.

\section{MSSM stop sector}
\label{sec:MSSM_stop}

In this study, we work in the MSSM and focus primarily on the third generation squark sector.  We decouple other SUSY particles: the gluino, sleptons, and the first and second generation squarks. We also decouple the non-SM Higgs particles by setting $M_A$ large.  The remaining SUSY particles in the model are the third generation squarks,  the neutralinos and charginos.

The gauge eigenstates for the superpartners of the top and bottom quarks are $(\tilde t_L,\tilde b_L), \tilde t_R$ and $\tilde b_R$, with  the left-handed states  grouped as an $ {\rm SU(2)}_L$ doublet and the right-handed states as singlets.   The mass matrix for the stop sector is
 \begin{equation}
  \bf{m_{\tilde t}^2} =
  \begin{pmatrix}
    M_{3SQ}^2 + m_t^2 + \Delta_{\tilde u_L} & m_t \tilde A_t \\
    m_t \tilde A_t & M_{3SU}^2 + m_t^2 + \Delta_{\tilde u_R}
  \end{pmatrix},
\end{equation}
with  $M_{3SQ}^2$ and $M_{3SU}^2$ representing the soft SUSY breaking masses for $\tilde{t}_L$ and $\tilde{t}_R$,   $m_t^2$ term coming from the F-term contribution in the SUSY Lagrangian and the $\Delta$ terms coming from the D-term contribution.
 \begin{comment}
\begin{equation}
  \Delta_{\tilde u_L} = \left(\frac{1}{2} - \frac{2}{3} \sin^2{\theta_W}\right) \cos{2\beta} ~m_Z^2
\end{equation}

\begin{equation}
  \Delta_{\tilde u_R} = \frac{2}{3} \sin^2{\theta_W} \cos{2 \beta} ~m_Z^2
\end{equation}
\end{comment}
The off-diagonal term $\tilde A_t$ is given by:
\begin{equation}
\tilde A_t = A_t - \mu/\tan{\beta},
\end{equation}
for $A_t$ representing the trilinear coupling and $\mu$ representing the supersymmetric bilinear mass term in the Higgs sector.   $\tan\beta=\langle H_u^0 \rangle/\langle H_d^0 \rangle$ is the ratio of the  vacuum expectation values of $H_u^0$ and $H_d^0$ in the MSSM.

The stop mass matrix can be diagonalized with a stop mixing angle $\theta_t$:
\begin{equation}
\begin{pmatrix}
\tilde{t}_1 \\ \tilde{t}_2
\end{pmatrix} =
	\begin{pmatrix}
	\cos{\theta_t} & -\sin{\theta_t}\\
	\sin{\theta_t} &  \cos{\theta_t}
	\end{pmatrix}
\begin{pmatrix}
\sL \\ \sR
\end{pmatrix},
\end{equation}
with mass eigenstates $\tilde{t}_1$, $\tilde{t}_2$: $m_{\tilde{t}_1} < m_{\tilde{t}_2}$.  For $M_{3SQ}< (>) M_{3SU}$, $\tilde{t}_1$ is mostly left-handed  (right-handed), while for $M_{3SQ}^2 \sim M_{3SU}^2$, $\tilde{t}_{1,2}$ could be a mixture of the left- and right-handed states.

Given the large top Yukawa coupling, the stop sector provides the dominant contribution to the radiative corrections of the SM-like Higgs mass in the MSSM.  For $M_{3SQ}=M_{3SU}=M_{SUSY}$, the correction to the SM-like Higgs mass squared is~\cite{Carena:1995bx}:
\begin{equation}
\delta m_h^2 = \frac{3}{4\pi^2} y_t^2 m_t^2 \sin^2{\beta}
	\left(	\log{\frac{M_{SUSY}^2}{m_t^2}} + \frac{\tilde A_t^2}{M_{SUSY}^2} \left( 1 - \frac{\tilde A_t^2}{12 M_{SUSY}^2}   \right)   \right).
\end{equation}
 In the minimal mixing case with $\tilde{A}_t=0$, a large $M_{SUSY}$ around 5$-$10 TeV is needed to provide a SM-like Higgs mass of 125 GeV.  In the maximal mixing case with $\tilde{A}_t=\sqrt{6}M_{SUSY}$, a relatively small $M_{SUSY} \sim$ TeV  can be accommodated given the additional contribution from the $\tilde{A}_t$ term.  In the general MSSM when  $M_{3SQ}^2\neq M_{3SU}^2$, to provide a SM-like Higgs mass of 125 GeV,    the light stop $\tilde{t}_1$ can still be as light as 200 GeV.  A large mass splitting between the stop mass eigenstates (and a large $\tilde{A}_t$ term), however,  is typically needed, resulting in $m_{\tilde{t}_2} \gtrsim 500$ GeV in general \cite{Christensen:2012ei, Carena:2011aa}.

Similarly,  the mass matrix for the sbottom  is given as:
\begin{equation}
  \bf{m_{\tilde b}^2} =
  \begin{pmatrix}
    M_{3SQ}^2 + m_b^2 + \Delta_{\tilde d_L} & m_b \tilde A_b \\
    m_b \tilde A_b & M_{3SD}^2 + m_b^2 + \Delta_{\tilde d_R}
  \end{pmatrix},
\end{equation}
with
\begin{equation}
\tilde A_b = A_b - \mu\tan{\beta}.
\end{equation}
Given the suppression of the off-diagonal terms  by the small bottom mass,   mixing among the sbottom mass eigenstates is typically small.  For $\tilde{A}_b\sim$ TeV, the sbottom mixing angle is about one degree for $M_{3SQ} \sim M_{3SD}\sim$ TeV. 

Since the stop sector provides the dominant contribution to the  Higgs mass corrections, we decouple the  right-handed sbottom in our analysis.  The left-handed sbottom mass, however,  is determined by $M_{3SQ}$ and could be relatively light.   Given $m_b \tilde A_b, M_{3SQ}^2 << M_{3SD}^2$, the light sbottom mass eigenstate is mostly left-handed:  $\tilde{b}_1 \sim \bL$.   Although the sbottom corrections to the Higgs mass are small compared to the stop corrections, there can be significant modifications to the Higgs couplings, especially  the bottom Yukawa coupling~\cite{Belyaev:2013rza}.

\section{Stop Decay}
\label{sec:prod_decay}

We consider the neutralino/chargino spectrum with a Bino-like LSP.   For simplicity, we consider three representative scenarios:
\begin{itemize}
\item{Case I}, Bino-like  LSP with decoupled Winos and Higgsinos:  $M_1< m_{\tilde{t},\tilde{b}_1} \ll |\mu|, M_2  $.
\item{Case IA}, Bino-like  LSP with Wino-like  NLSPs: $M_1 < M_2 <  m_{\tilde{t},\tilde{b}_1} \ll |\mu|$.
\item{Case IB}, Bino-like  LSP with Higgsino-like  NLSPs: $M_1< |\mu|<   m_{\tilde{t},\tilde{b}_1} \ll M_2$.
\end{itemize}
The decays of the light stop or sbottom highly depend on
the low-lying neutralino/chargino spectrum, as well as the composition of the light stop and sbottom.

In each scenario, we consider two limiting cases with different stop left-right mixing.  In the minimal mixing case, $\tilde{A}_t=A_t - \mu\cot\beta=0$,   the lightest stop mass eigenstate $\tilde{t}_1$ is  either purely $\tilde{t}_L$ ($M_{3SQ} < M_{3SU}$) or purely $\tilde{t}_R$ ($M_{3SQ} > M_{3SU}$).    We decouple  $\tilde{t}_2$  for simplicity.    In the  maximal mixing case with  $M_{3SQ}=M_{3SU}=M_{SUSY}$ and $|\tilde{A}_t|= \sqrt{6} M_{SUSY}$,   both $\tilde{t}_{1,2}$ are  a mixture of $\tilde{t}_L$ and $\tilde{t}_R$,  with mass squared splitting $\Delta m_{\tilde t}^2 \approx 2 \sqrt{6} m_t M_{SUSY}$.   In our analysis below, we use $\tilde{A}_t>0$.  Negative values of $\tilde{A}_t$ introduce little changes to the numerical results.  Since $M_{3SQ} $ also controls the mass for $\tilde{b}_L$, there is   a light $\tilde{b}_1\sim \tilde{b}_L$ for the light $M_{3SQ}$ case,  assuming small sbottom left-right mixing and a decoupled $\tilde{b}_R$.

\begin{figure}
\begin{center}
\includegraphics[width = 0.45\textwidth]{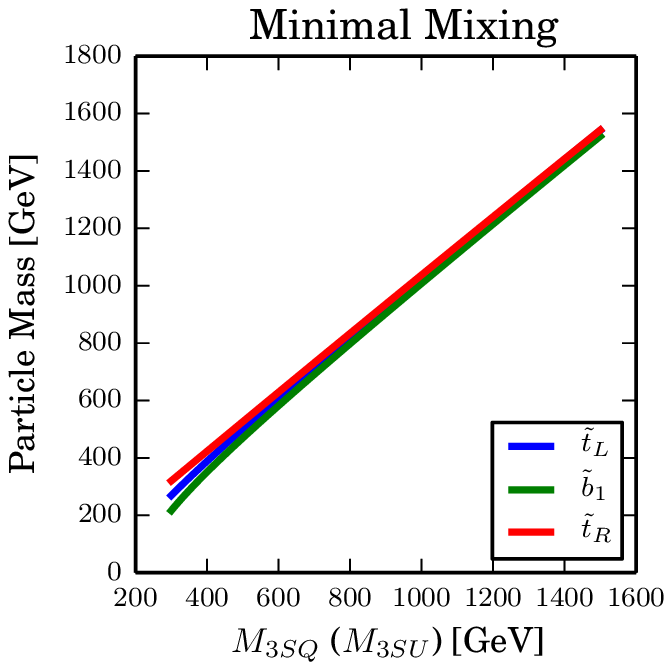}
\includegraphics[width = 0.45\textwidth]{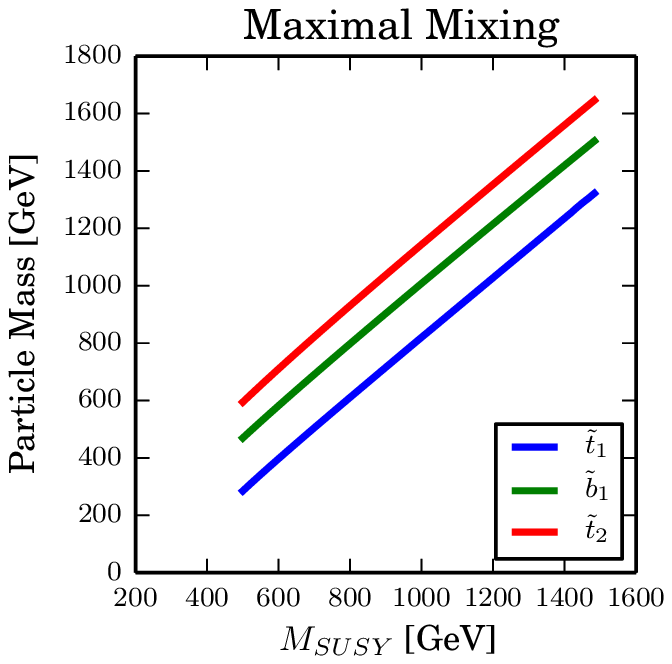}
\end{center}
\caption{The mass spectra for stops and sbottom for the minimal mixing case (left panel) and the maximal mixing case with  $M_{3SQ}=M_{3SU}=M_{SUSY}$ (right panel).}
\label{Figure:mass}
\end{figure}

The mass spectra for stops and sbottom are shown in Fig.~\ref{Figure:mass}.   In the minimal mixing case (left panel), $m_{\tilde{t}_L}$ ($m_{\tilde{t}_R}$), $m_{\tilde{b}_1}$ $\sim M_{3SQ} (M_{3SU})$, especially for large $M_{3SQ}$ ($M_{3SU}$).   In the maximal mixing case (right panel), the mass difference between $\tilde{b}_1$ and $\tilde{t}_1$ is typically about 250 GeV while the mass difference between $\tilde{t}_2$ and $\tilde{t}_1$ is about 350 GeV or larger.

We used SUSY-HIT \cite{Djouadi:2006bz} to calculate the supersymmetric particle spectrum and decay branching fractions.  In this section, unless otherwise specified,  we have set the Bino-like LSP  mass parameter $M_1=150$ GeV,     the intermediate gaugino mass parameters   $M_2,\mu = 300$ GeV  in Cases IA and IB,  respectively, and $\tan\beta=10$.

\subsection{Case I:  Bino-like LSP with decoupled Wino and Higgsino}

The simplest case has a mass spectrum with stop(s), left-handed sbottom, and   only the low-lying neutralino being the Bino-like LSP.

In the minimal mixing case with  the light stop $\tilde{t}_1$ as a pure left- or right-handed state, $\tilde{t}_1$ either directly decays to $t \chi_1^0$ when it is kinematically accessible or through $bW^{*} \chi_1^0$ with 100\% branching fraction.  Similarly, in the case of small $M_{3SQ}$, $\tilde{b}_1$ decays directly through $b \chi_1^0$ with 100\% branching fraction.

\begin{figure}
\begin{center}
\includegraphics[width =  \textwidth]{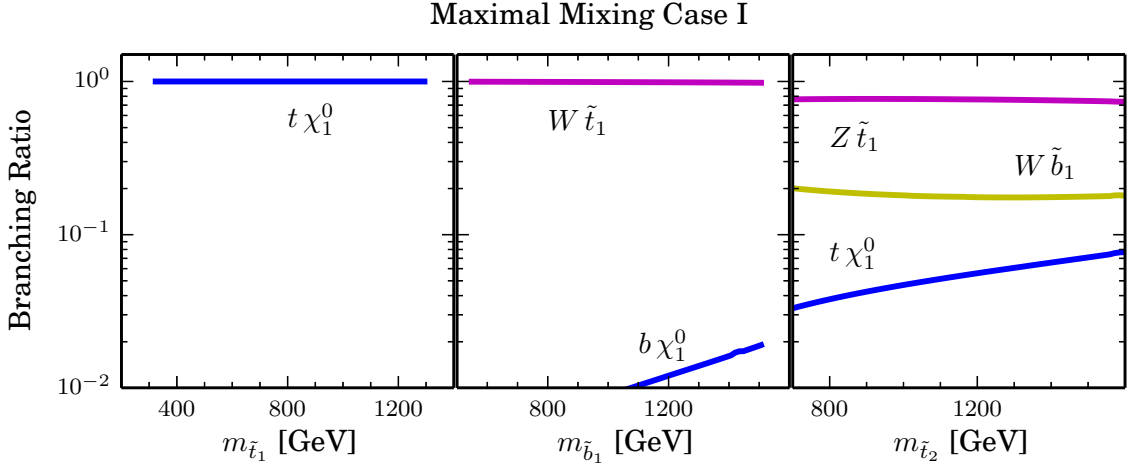}
 \end{center}
\caption{Branching fractions for $\tilde{t}_1$ (left), $\tilde{b}_1$ (middle) and $\tilde{t}_2$ (right) in the  maximal mixing scenario with a Bino-like LSP (Case I).    We set $M_1=$ 150 GeV, $M_2=$ 2 TeV,  $\mu=$  2 TeV, and $\tan\beta=10$, which corresponds to   $m_{\chi_1^0}=151$ GeV.}
\label{Figure:simple_max}
\end{figure}

In the maximal mixing case,   $\tilde{t}_1$, $\tilde{t}_2$, and $\tilde{b}_1$ appear in the spectrum, with a typical mass order  $m_{\tilde{t}_1} <  m_{\tilde{b}_1} < m_{\tilde{t}_2}$ with relatively large mass splittings of 150 GeV or larger.    While the decay of  $\tilde{t}_1$ is straightforward (100\% into $bW^{(*)}\chi_1^0$), the decays of $\tilde{b}_1$ and $\tilde{t}_2$ could have multiple competing channels, as shown in Fig.~\ref{Figure:simple_max}.   For $\tilde{b}_1$, it dominantly decays into $W \tilde{t}_1$ while the branching fraction of the  $\tilde{b}_1 \rightarrow b \chi_1^0$ channel is only about a few percent or less.   For   $\tilde{t}_2$, it dominantly decays into a light stop/sbottom with a gauge boson:  $Z \tilde{t}_1 $ about 75\% and $W \tilde{b}_1$ about 20\%.  The direct decay down to $t \chi_1^0$ is less than   10\%.

\begin{figure}
\begin{center}
\includegraphics[width = \textwidth]{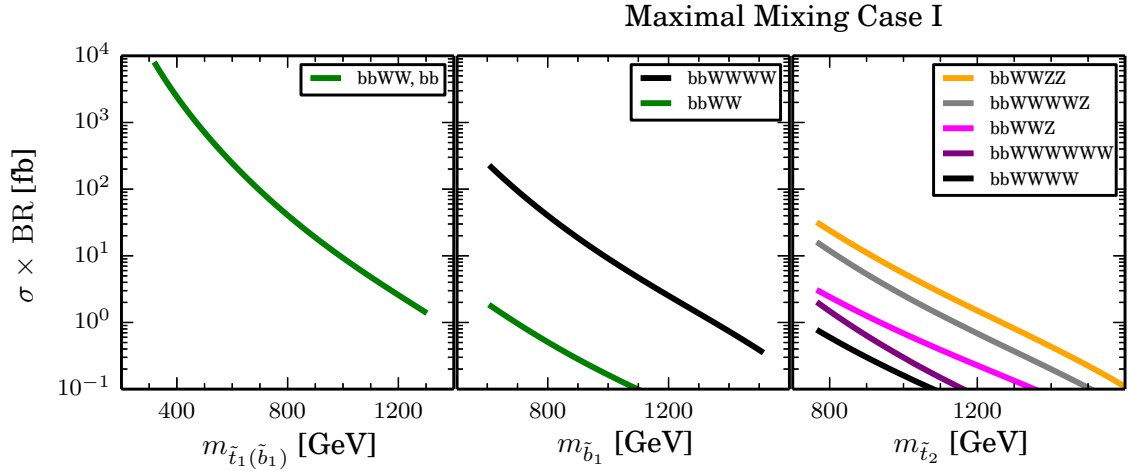}
\end{center}
\caption{Case I: left panel shows $\sigma\times{\rm BR}$ of  final states for $\tilde{t}_1$ pair production in both the minimal and maximal mixing scenarios, as well as $\tilde{b}_1$ pair production in the minimal mixing scenario.   The middle and right panel show  $\sigma\times{\rm BR}$ for various final states of $\tilde{b}_1$ and $\tilde{t}_2$ pair production,  respectively,  in the maximal mixing scenario.   All channels include $\met$ in the final states.    All the cross sections are for  the 14 TeV LHC stop and sbottom pair production, calculated including NLO + NLL corrections~\cite{oai:arXiv.org:1006.4771, Broggio:2013uba,Borschensky:2014cia}. The choice of neutralino and chargino mass parameters is the same as in Fig.~\ref{Figure:simple_max}.}
\label{Figure:CSBR_caseI}
\end{figure}

The pair production of stops and sbottoms at the LHC, and their subsequent decays result in the appearance  of various final states.   In the left panel of Fig.~\ref{Figure:CSBR_caseI}, we show the $\sigma\times{\rm BR}$ of  final states $tt/bbWW+\met$ for $\tilde{t}_1$ in the  minimal   and maximal mixing scenarios, as well as $bb+\met$ for $\tilde{b}_1$ in the  minimal mixing scenario  at the 14 TeV LHC.     All the cross sections shown in the plots are for stop and sbottom pair production at 14 TeV including NLO supersymmetric QCD correction as well as resummation of soft-gluon emission at next-to-leading logarithmic accuracy~\cite{oai:arXiv.org:1006.4771,Broggio:2013uba,Borschensky:2014cia}.    Since  $\tilde{t}_1 \rightarrow t/bW \chi_1^0$ and $\tilde{b}_1 \rightarrow b\chi_1^0$ dominate in those channels,     $\sigma\times{\rm BR}$ is the same as the production cross sections for the stop pair and sbottom pair.    The middle panel of Fig.~\ref{Figure:CSBR_caseI} shows the $\sigma\times{\rm BR}$ for $\tilde{b}_1\tilde{b}_1$ pair production in the maximal mixing scenario.  The    $bb+\met$ channel is highly suppressed, while $bbWWWW+\met$ becomes dominant.      The right panel of Fig.~\ref{Figure:CSBR_caseI} shows the $\sigma\times{\rm BR}$ for $\tilde{t}_2\tilde{t}_2$ pair production in the maximal mixing scenario.   The  dominant channel is $ttZZ+\met$,  with   $ttWWZ$ being the second dominant channel.  The cross section, however, is relatively small, less than about 10 fb for $m_{\tilde{t}_2} \gtrsim 800$ GeV, given the heaviness of the   second stop.    Note that the range of the stop and sbottom masses are controlled by the choice of parameter $M_{3SQ}=M_{3SU}=M_{SUSY}=600 \ldots 1500 $ GeV in the maximal mixing case(see Fig.~\ref{Figure:mass}).

\subsection{Case IA: Bino LSP with Wino NLSP}

\begin{figure}[t]
\begin{center}
\includegraphics[width =  \textwidth]{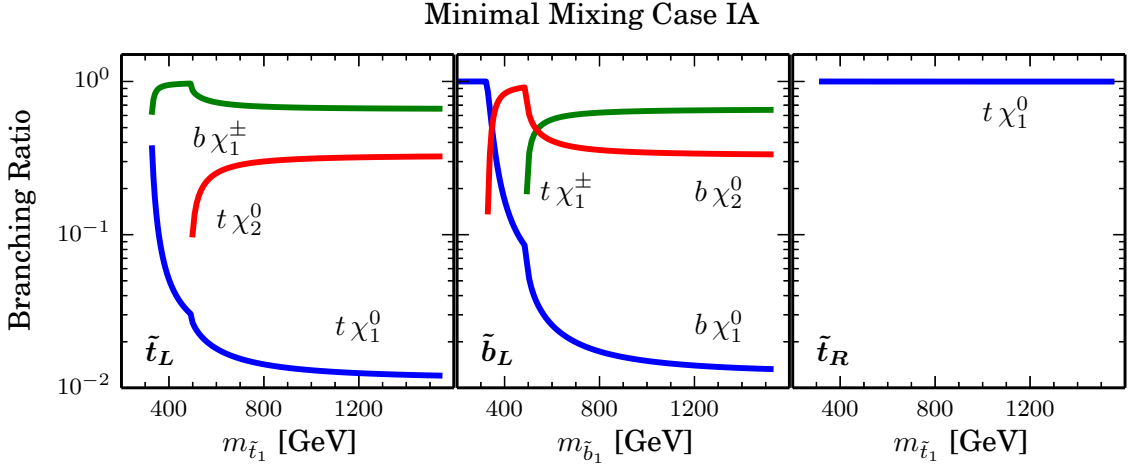}
 \end{center}
\caption{Case IA:  branching fractions for left-handed $\tilde{t}_1$ (left), $\tilde{b}_1$ (middle), right-handed $\tilde{t}_1$ (right)  in the  minimal mixing scenario.    We set $M_1=$ 150 GeV, $M_2=$ 300 GeV,  $\mu=$  2 TeV, and $\tan\beta=10$, which corresponds to   $m_{\chi_1^0}=151$ GeV, $m_{\chi_2^0}=319$ GeV and  $m_{\chi_1^\pm}=319$ GeV.} 
\label{Figure:M2_left}
\end{figure}

The low lying  neutralino/chargino spectrum in Case IA comprises of a Bino-like LSP, as well as a pair of Wino-like states: $\chi_2^0$ and $\chi_1^\pm$ with nearly degenerate masses.   In the minimal mixing scenario, the decay branching fractions are shown in Fig.~\ref{Figure:M2_left} for   left-handed $\tilde{t}_1$ (left), $\tilde{b}_1$ (middle), and right-handed $\tilde{t}_1$ (right).   For the left-handed $\tilde{t}_1$, decays to $b\chi_1^\pm$ ($\sim$ 70\% for large $m_{\tilde{t}_1}$) and $t\chi_2^0$ ($\sim$ 30\% for large $m_{\tilde{t}_1}$) dominate over $t\chi_1^0$ once    kinematically accessible, due to the stronger  ${\rm SU}(2)_L$ coupling compared to   the relatively weaker   ${\rm U}(1)_Y$ coupling.   Similarly, $\tilde{b}_1 \rightarrow t \chi_1^\pm$ ($\sim$ 65\%) and $\tilde{b}_1 \rightarrow b \chi_2^0$ ($\sim$ 30\%) dominate over the      $b \chi_1^0$ channel for sbottom.    Given the dominant decay channels of the  Wino-like neutralino/chargino\footnote{For $\chi_2^0$, whether it decays preferably to $Z\chi_1^0$ or $h\chi_1^0$ depends on the sign of $\mu$, as explained in detail in Ref.~\cite{Han:2013kza}.}:  $\chi_1^\pm \rightarrow W \chi_1^0$,  $\chi_2^0 \rightarrow Z/h \chi_1^0$,  the dominant decay modes for $\tilde{t}_1$ and $\tilde{b}_1$ are:  $\tilde{t}_1 \rightarrow bW\chi_1^0, \ tZ/h \chi_1^0$, $\tilde{b}_1 \rightarrow tW\chi_1^0, \ bZ/h \chi_1^0$.   When $\tilde{t}_1$ is mostly right-handed, it decays to $t\chi_1^0$ almost 100\%, since its couplings to the Wino-like neutralino/charginos are highly suppressed.

\begin{figure}
\begin{center}
\includegraphics[width =  \textwidth]{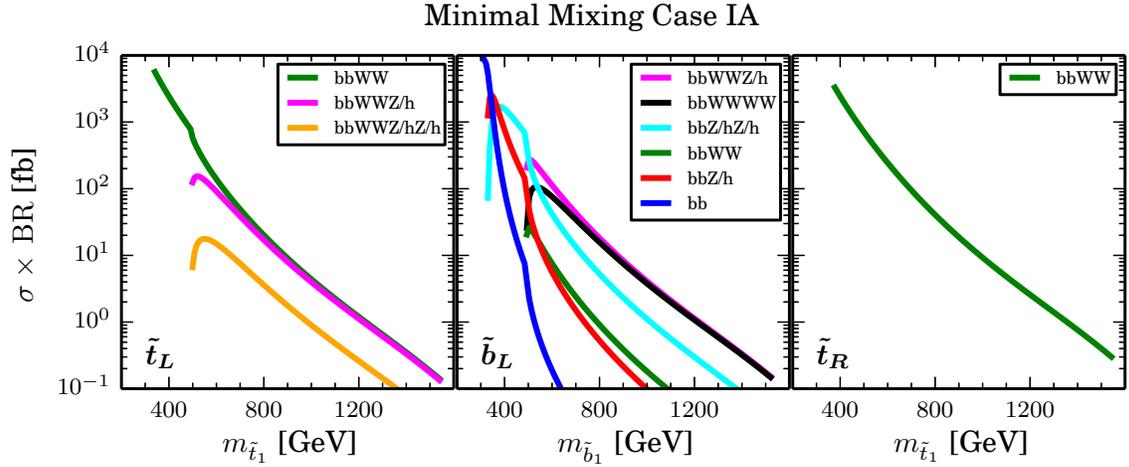}
  \end{center}
\caption{Case IA: $\sigma\times{\rm BR}$ of various final states for pair production of left-handed $\tilde{t}_1$ (left), $\tilde{b}_1$ (middle), and right-handed $\tilde{t}_1$ (right) in the minimal mixing scenario at the 14 TeV LHC.      The choice of neutralino and chargino  mass parameters is the same as in Fig.~\ref{Figure:M2_left}. }    
\label{Figure:CSBR_caseIA_min}
\end{figure}

The left, middle and right panels of Fig.~\ref{Figure:CSBR_caseIA_min} show the $\sigma \times {\rm BR}$ for pure left-handed $\tilde{t}_1\tilde{t}_1$, $\tilde{b}_1\tilde{b}_1$  and pure right-handed $\tilde{t}_1\tilde{t}_1$ pair production, respectively,  in the minimal mixing scenario of Case IA.   For pure left-handed $\tilde{t}_1$,  $bbWWZ/h+\met$ is as  abundant  as the   $bbWW+\met$ channel, which could be an important new search channel for the stop. 
For pure left-handed $\tilde{b}_1$,  the  $bb+\met$ channel is highly suppressed.  New final states of $bbWWZ/h$ and  $bbWWWW$ are dominant and comparable in size, with $bbZ/hZ/h$ being subdominant,  opening up new channels for sbottom searches.    The final state for the pure right-handed $\tilde{t}_1$ is still $bbWW+\met$, despite the existence of  light Wino NLSPs in the spectrum. 

\begin{figure}[t]
\begin{center}
\includegraphics[width = \textwidth]{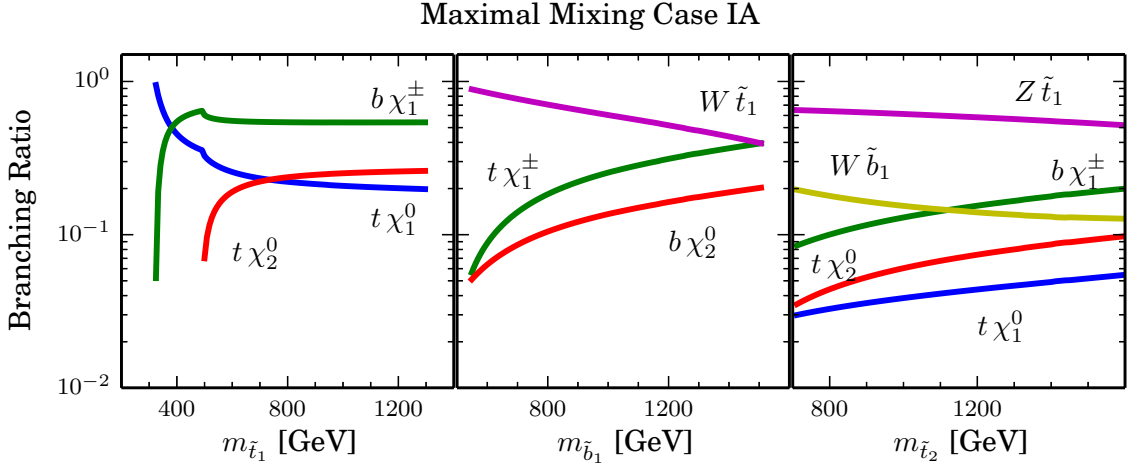}
 \end{center}
\caption{Case IA: Branching fractions for $\tilde{t}_1$ (left), $\tilde{b}_1$ (middle) and $\tilde{t}_2$ (right) in the maximal mixing scenario.    The choice of neutralino and chargino  mass parameters is the same as in Fig.~\ref{Figure:M2_left}. }   
\label{Figure:M2_max}
\end{figure}

For the maximally mixed scenario, the decay of $\tilde{t}_1$,  $\tilde{b}_1$ and $\tilde{t}_2$ are shown in the left, middle and right panels of Fig.~\ref{Figure:M2_max},  respectively.  For $\tilde{t}_1$ with large mass,   the decay to $b\chi_1^\pm$, $t\chi_2^0$ still dominates over $t\chi_1^0$,  but  the corresponding branching fractions are smaller compared to the pure left-handed case (Fig.~\ref{Figure:M2_left}) due to the decrease of the coupling to the Wino-like state caused by the right-handed stop component.    For $\tilde{b}_1$,  while $t\chi_1^\pm$ and $b\chi_2^0$ modes still dominate over $b\chi_1^0$ mode,  the new decay channel of $W\tilde{t}_1$ opens up and even dominates over most of the mass range.    Its branching fraction varies between 100\%  to  about 40\% for $m_{\tilde{b}_1}$ between 600 GeV to 1500 GeV.    For $\tilde{t}_2$, in addition to $b\chi_1^\pm$ and $t \chi_{1,2}^0$ (about a few percent to 20\%), decays to a light stop/sbottom  plus a gauge boson~\cite{Ghosh:2013qga} become comparable or even dominant: about 50\% $-$ 70\% for $Z\tilde{t}_1$ and about 20\% $-$ 15\% for $W\tilde{b}_1$.

\begin{figure}
\begin{center}
\includegraphics[width =  \textwidth]{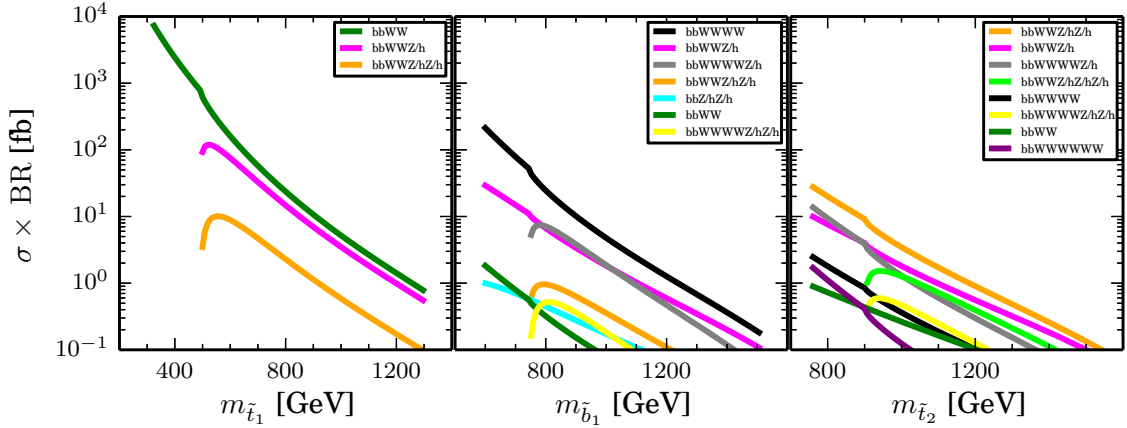}
  \end{center}
\caption{Case IA: $\sigma\times{\rm BR}$ of various final states for pair production of  $\tilde{t}_1$ (left), $\tilde{b}_1$ (middle), and  $\tilde{t}_1$ (right) in the maximal mixing scenario at the 14 TeV LHC.     The choice of neutralino and chargino  mass parameters is the same as in Fig.~\ref{Figure:M2_left}. }  
\label{Figure:CSBR_caseIA_max}
\end{figure}

The left, middle and right panels of Fig.~\ref{Figure:CSBR_caseIA_max} show the $\sigma \times {\rm BR}$ for $\tilde{t}_1\tilde{t}_1$, $\tilde{b}_1\tilde{b}_1$, and $\tilde{t}_2\tilde{t}_2$ respectively for the maximal mixing scenario of Case IA at the 14 TeV LHC.    For the light stop, while the dominant channel is  still  $bbWW+\met$, the subdominant channel $bbWWZ/h+\met$ could still have a sizable cross section.     For the light sbottom, $bbWWWW+\met$ becomes dominant.  For the heavy stop, multiple channels open, with $bbWWZ/h Z/h+\met$ being dominant, followed by   $bbWWZ/h+\met$, $bbWWWWZ/h +\met$, and    $bbWWZ/hZ/hZ/h +\met$.

\subsection{Case IB: Bino-LSP with Higgsino-NLSP}

 \begin{figure}[t]
\begin{center}
\includegraphics[width =  \textwidth]{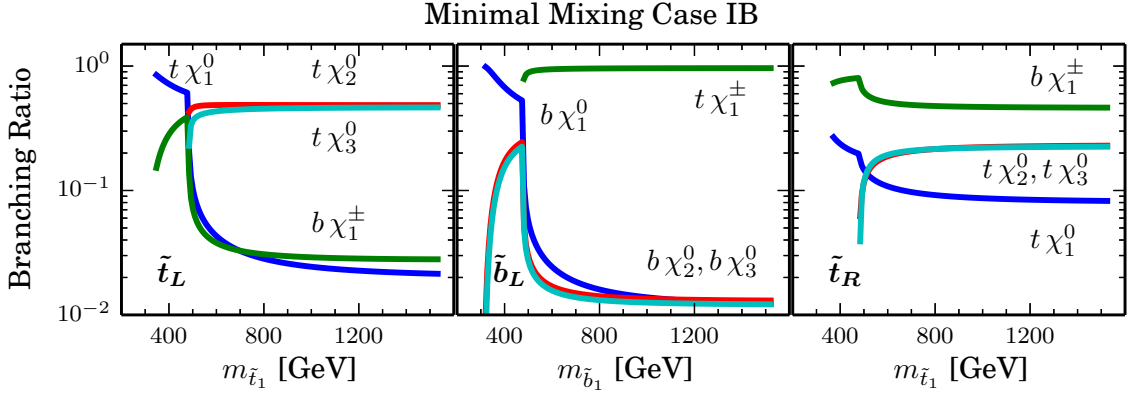}
 \end{center}
\caption{Case IB:  branching fractions for left-handed $\tilde{t}_1$ (left), $\tilde{b}_1$ (middle), right-handed $\tilde{t}_1$ (right)  in the  minimal mixing scenario.     We set $M_1=$ 150 GeV, $\mu=$ 300 GeV,  $M_2=$  2 TeV, and $\tan\beta=10$, which corresponds to   $m_{\chi_1^0}=145$ GeV, $m_{\chi_2^0}=308$ GeV, $m_{\chi_3^0}=310$ GeV and  $m_{\chi_1^\pm}=304$ GeV.} 
\label{Figure:mu_min}
\end{figure}

The low lying  neutralino/chargino spectrum in Case IB comprises of a Bino-like LSP, as well a pair of Higgsino-like neutralino states $\chi_{2,3}^0$ and chargino states $\chi_1^\pm$ with nearly degenerate masses.   Fig.~\ref{Figure:mu_min} shows the branching fractions of left-handed $\tilde{t}_1$ and $\tilde{b}_1$ and right-handed $\tilde{t}_1$ in the left, middle and right panels for the minimal mixing scenario.   For $\tilde{t}_1$, decays to $t\chi_{2,3}^0$ dominate over $b\chi_1^\pm$ and $t\chi_1^0$ since the former ones are controlled by the large top Yukawa coupling, compared to the small bottom Yukawa coupling and ${\rm U}(1)_Y$ couplings for the latter two.    However, for $\tilde{b}_1$, the decay of $t\chi_1^\pm$ becomes dominant  since the $\tilde{b}_L\bar{t}_R\tilde{H}_u^+$ coupling is proportional to the top Yukawa while its couplings to $\chi_{2,3}^0$ and $\chi_1^0$ are suppressed by the bottom Yukawa coupling and ${\rm U}(1)_Y$ couplings.    For the right-handed $\tilde{t}_1$ case, it dominantly decays to $b\chi_1^\pm$, reaching almost 50\%, while decays to $t\chi_{2}^0+t\chi_3^0$ are about 20\%.  All channels are controlled by the top Yukawa coupling while the latter ones have extra phase space suppression.   Given the near degeneracy of the two Higgsino states $\chi_{2,3}^0$,  contributions from final states involving $\chi_{2,3}^0$ are usually summed over in collider analyses.

\begin{figure}
\begin{center}
\includegraphics[width =  \textwidth]{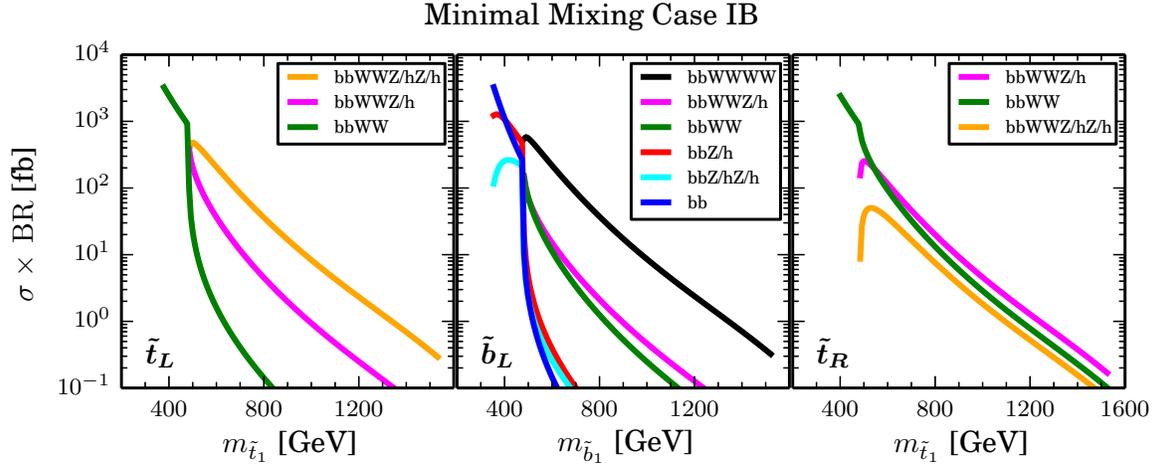}
  \end{center}
\caption{Case IB: $\sigma\times{\rm BR}$ of various final states for pair production of left-handed $\tilde{t}_1$ (left), $\tilde{b}_1$ (middle), and right-handed $\tilde{t}_1$ (right) in the minimal mixing scenario at the 14 TeV LHC.      The choice of neutralino and chargino  mass parameters is the same as in Fig.~\ref{Figure:mu_min}. } 
\label{Figure:CSBR_caseIB_min}
\end{figure}

Given the further decays of $\chi_1^\pm \rightarrow W \chi_1^0$, $\chi_{2,3}^0 \rightarrow Z \chi_1^0/h\chi_1^0$ as discussed in detail in~\cite{Han:2013kza}, the pair production of stops and sbottoms lead to complicated final states at the collider.  The left, middle and right panels of Fig.~\ref{Figure:CSBR_caseIB_min} show the $\sigma \times {\rm BR}$ for pure left-handed $\tilde{t}_1\tilde{t}_1$, $\tilde{b}_1\tilde{b}_1$  and pure right-handed $\tilde{t}_1\tilde{t}_1$ pair production in the minimal mixing scenarios of Case IB.   For pure left-handed $\tilde{t}_1$,  $bbWWZ/hZ/h+\met$ is the dominant final state with the  stop search channel $bbWW+\met$ being highly suppressed.
For pure left-handed $\tilde{b}_1$,   $bbWWWW+\met$ is  the dominant channel.    The  dominant final states for pure right-handed $\tilde{t}_1$ are  $bbWWZ/h+\met$ as well as  $bbWW+\met$.

  \begin{figure}[t]
\begin{center}
\includegraphics[width =  \textwidth]{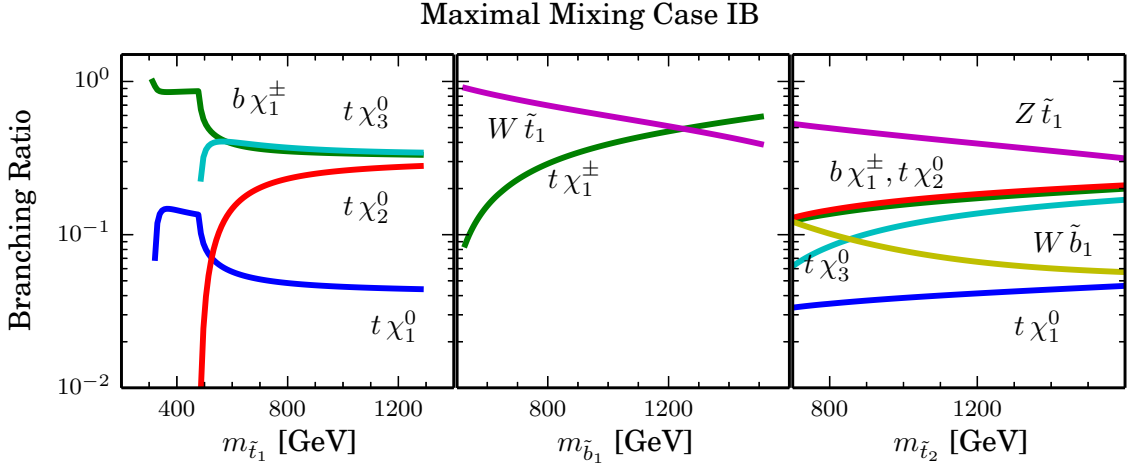}
 \end{center}
\caption{Case IB: Branching fractions for $\tilde{t}_1$ (left), $\tilde{b}_1$ (middle) and $\tilde{t}_2$ (right) in the maximal mixing scenario.   The choice of neutralino and chargino  mass parameters is the same as in Fig.~\ref{Figure:mu_min}. } 
\label{Figure:mu_max}
\end{figure}

For the maximal mixing scenario, the decay branching fractions for $\tilde{t}_1$, $\tilde{b}_1$, and $\tilde{t}_2$ are shown in the left, middle and right panels of Fig.~\ref{Figure:mu_max}, respectively.   Since $\tilde{t}_1$ is an equal mixture of left- and right-handed components, the decays to $t\chi_{2,3}^0$ (dominant for $\tilde{t}_L$) and $b\chi_1^\pm$ (dominant for $\tilde{t}_R$)  (see the left and right panel of Fig.~\ref{Figure:mu_min}) have roughly the same decay branching fraction,  around 30\% each.  Decay to the final state of $t\chi_1^0$ is typically a few percent, unless other decay modes are kinematically unaccessible at small $m_{\tilde{t}_1}$.

For $\tilde{b}_1$,   the relative strength of $t\chi_1^\pm$ and $b\chi_{2,3}^0$ is similar to that of the $\tilde{b}_1$ in the minimal mixing scenario,  but the opening of the $W\tilde{t}_1$ mode dominates the decay for most of the mass range, leading to the suppression of the  $t\chi_1^\pm$ and $b\chi_{2,3}^0$ modes.  With increasing $m_{\tilde{b}_1}$,  $t\chi_1^\pm$ becomes more and more important, which dominates over $W\tilde{t}_1$ when $m_{\tilde{b}_1} \gtrsim$ 1200 GeV.

For $\tilde{t}_2$, decay to $Z\tilde{t}_1$ is dominant, about 60\% $-$ 30\% for $m_{\tilde{t}_2}$ in the range of 700 $-$ 1600 GeV.   Decays to $b\chi_1^\pm$, $t\chi_{2,3}^0$ are sub-dominant, around 10\% $-$ 20\% for each channel.  $\tilde{t}_2 \rightarrow W \tilde{b}_1$ is typically around 10\% to about a few percent, while $\tilde{t}_2 \rightarrow t \chi_1^0$ is only at a few percent level.

\begin{figure}
\begin{center}
\includegraphics[width =  \textwidth]{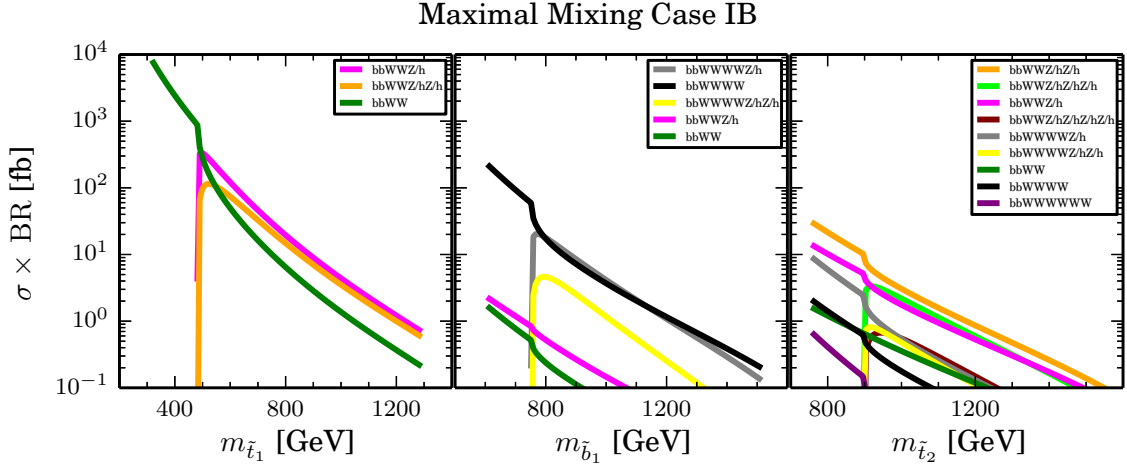}
  \end{center}
\caption{Case IB: $\sigma\times{\rm BR}$ of various final states for pair production of  $\tilde{t}_1$ (left), $\tilde{b}_1$ (middle), and  $\tilde{t}_2$ (right) in the maximal mixing scenario at the 14 TeV LHC.   The choice of neutralino and chargino  mass parameters is the same as in Fig.~\ref{Figure:mu_min}.   }
\label{Figure:CSBR_caseIB_max}
\end{figure}

The left, middle and right panel of Fig.~\ref{Figure:CSBR_caseIB_max} show the $\sigma \times {\rm BR}$ for $\tilde{t}_1\tilde{t}_1$, $\tilde{b}_1\tilde{b}_1$, and $\tilde{t}_2\tilde{t}_2$ for the maximal mixing scenario of Case IB at the 14 TeV LHC.    For the light stop,  the dominant channel is  $bbWWZ/h+\met$, followed by $bbWWZ/hZ/h+\met$.    The   $bbWW+\met$ channel is suppressed by about a factor of 5.   For the light sbottom, $bbWWWW+\met$ and $bbWWWWZ/h+\met$ are dominant.  For the heavy stop, multiple channels open, with $bbWWZ/hZ/h+\met$ being dominant, followed by  $bbWWZ/hZ/hZ/h +\met$ and  $bbWWZ/h+\met$.

%%%%

\section{Current  collider search limits on stop and sbottom}
\label{sec:limit}

Searches for direct stop and sbottom pair production have been performed at both ATLAS and CMS, with about 20 ${\rm fb}^{-1}$ data  at $\sqrt{s}=8$ TeV, and about 5 ${\rm fb}^{-1}$ data at $\sqrt{s}=7$ TeV~\cite{Aad:2014bva,Aad:2014kra,Aad:2014qaa,CMS:2013nia,CMS-PAS-SUS-14-011,Chatrchyan:2013xna,Aad:2014nra, CMS-PAS-SUS-13-009,Aad:2014mha,Khachatryan:2014doa,CMS-PAS-SUS-13-021,Chatrchyan:2013mya,Aad:2013ija,CMS-PAS-SUS-13-018,Aad:2014lra,CMS-PAS-SUS-13-008,Aad:2014pda,Chatrchyan:2013fea,Chatrchyan:2014aea}.  Here we summarize the current experimental search channels and exclusion bounds assuming a stable neutralino LSP.  Stop searches   in scenarios  with  a Gravitino LSP have been analyzed in Refs.~\cite{Aad:2014mha, Chatrchyan:2013mya}.

\begin{itemize}
\item{$\tilde{t}_1 \to t^{(*)}  \chi_1^0$ and $\tilde{t}_1 \to b \chi_1^\pm \to bW^{(*)} \chi_1^0$~\cite{Aad:2014bva,Aad:2014kra,Aad:2014qaa,  CMS:2013nia, CMS-PAS-SUS-14-011,Chatrchyan:2013xna}}

ATLAS results on fully hadronic final states~\cite{Aad:2014bva} exclude stops in the regions of 270 $< m_{\tilde{t}_1} <$ 645 GeV for $m_{\tilde{\chi}_1^0} <$ 30 GeV, assuming both stops decay 100\% via $\tilde{t}_1 \to t \chi_1^0$.   Regions of 245 $< m_{\tilde{t}_1} <$ 400 GeV for $m_{\tilde{\chi}_1^0} <$ 60 GeV, $m_{\chi_1^\pm}=2 m_{\chi_1^0}$  are excluded when both stops  decay 100\% via $\tilde{t}_1 \to bW^{(*)}  \chi_1^0$.     For ${\rm BR}(\tilde{t}_1 \to bW^{(*)}  \chi_1^0)={\rm BR}(\tilde{t}_1 \to t \chi_1^0)=50\%$, stop masses in the range of 250 $-$ 550 GeV are excluded for $m_{\tilde{\chi}_1^0} <$ 60 GeV, $m_{\chi_1^\pm}=2 m_{\chi_1^0}$.

For semileptonic channels, stop masses between 210 GeV and 640 GeV are excluded at 95\% C.L. for a massless LSP, and stop masses around 550 GeV are excluded for LSP mass below 230 GeV~\cite{Aad:2014kra},  assuming ${\rm BR}(\tilde{t}_1 \to t \chi_1^0)=$100\%.   For ${\rm BR}(\tilde{t}_1 \to b \chi_1^\pm \rightarrow bW^{(*)}\chi_1^0)=$100\%, the excluded stop and LSP masses depend strongly on the mass of the $\chi_1^\pm$.    For $m_{\chi_1^\pm}=2m_{\chi_1^0}$, stop masses up to 500 GeV are excluded for   LSP masses in the range of 100 and 150 GeV.  
 For the compressed spectrum case when $m_{\chi_1^\pm} -m_{\chi_1^0}$ is small with soft leptons from $\chi_1^\pm$ decay, stop masses between 265 (240) GeV and 600 GeV are excluded for $m_{\chi_1^\pm} -m_{\chi_1^0}=5 (20)$ GeV with  an  LSP mass of 100 GeV.  For small mass splitting between $\tilde{t}_1$ and $\chi_1^\pm$ (for example, 10 GeV) with soft $b$ jets,  stop masses below 390 GeV are excluded for a massless LSP.  When both decay modes $\tilde{t}_1 \to t \chi_1^0$ and $\tilde{t}_1 \to b \chi_1^\pm$ are open,  the excluded stop masses increase from 530 GeV to 660 GeV for an LSP mass of 100 GeV when ${\rm BR}(\tilde{t}_1 \to t \chi_1^0)$ is increased from 0\% to 100\% and  $m_{\chi_1^\pm}=2m_{\chi_1^0}$.  The limits get weaker with an increased branching ratio to decays other than $\tilde{t}_1 \to t \chi_1^0$ and $\tilde{t}_1 \to b \chi_1^\pm$.

Limits from the pure leptonic channels are  weaker~\cite{Aad:2014qaa}.    Stops with masses between 215 GeV and 530  GeV decaying to an on-shell $t$-quark and a neutralino are excluded at 95\% C. L.  for a 1 GeV neutralino.  For $m_b+m_W+m_{\chi_1^0}< m_{\tilde{t}_1}<m_t+m_{\chi_1^0}$ with an off-shell top and a neutralino LSP, the stop masses are excluded between 90 GeV and 170 GeV.    For ${\rm BR}(\tilde{t}_1\rightarrow b \chi_1^\pm)=100\%$,  the limits on the stop mass depend on both the LSP mass and $m_{\chi_1^\pm}$.   $m_{\tilde{t}_1}$ between 150 GeV and 445 GeV  is excluded at 95\% C. L.  for $m_{\tilde{t}_1}=m_{\chi_1^\pm}+ 10$ GeV, in the case of a 1 GeV  neutralino LSP.    For $m_{\tilde{t}_1}=2m_{\chi_1^\pm}$, stop masses between 210 GeV and 340 GeV are excluded for an LSP mass of 100 GeV.   For a fixed $m_{\chi_1^\pm}=106$ GeV, stop masses between 240 GeV to 325 GeV are excluded with an LSP mass of 1 GeV.

Limits from CMS are very similar \cite{CMS:2013nia,CMS-PAS-SUS-14-011,Chatrchyan:2013xna}.  Note that limits on the stop exclusion depend on the branching fractions of $\tilde{t}_1 \to t^{(*)} \chi_1^0$  and $\tilde{t}_1\to b \chi_1^\pm$.   For $\tilde{t}_1 \to b \chi_1^\pm$, the limits also depend on the mass of the intermediate chargino.

\item{$\tilde{t}_1\rightarrow c \chi_1^0$ or $\tilde{t}_1\rightarrow b f f^\prime \chi_1^0$~\cite{Aad:2014nra,CMS-PAS-SUS-13-009}}

For small mass splitting between $m_{\tilde{t}_1}$ and $m_{\chi_1^0}$, stop decays via $\tilde{t}_1\rightarrow c \chi_1^0$ or $\tilde{t}_1\rightarrow b f f^\prime \chi_1^0$~\cite{Aad:2014nra,CMS-PAS-SUS-13-009}.   For 100\% branching fraction of $\tilde{t}_1\rightarrow c \chi_1^0$, searches on charm tagged events and monojet-like events exclude stop masses around 240 GeV for $\Delta m=m_{\tilde{t}_1}-m_{\chi_1^0}<85$ GeV.   Stop masses up to 270 GeV are excluded for an LSP mass of 200 GeV.  For nearly degenerate stop and LSP, stop masses up to about 260 GeV are excluded.   For 100\% branching fraction of $\tilde{t}_1\rightarrow b f f^\prime \chi_1^0$, searches based on monojet plus $\met$  exclude stop masses up to about 255 GeV for $\Delta m \sim m_b$ and about 150 (200) GeV for $m_b < \Delta m < 50 (35)$ GeV. 

For small mass splitting between $m_{\chi_1^\pm}$ and $m_{\chi_1^0}$ with undetectable decay products of $\chi_1^\pm$,  pair production of stop with $\tilde{t}_1\rightarrow b \chi_1^\pm$ leads to two $b$ jets plus $\met$ events.  Results from ATLAS~\cite{Aad:2013ija} exclude stop masses up to 580 (440) GeV for $m_{\chi_1^\pm}-m_{\chi_1^0}=5 (20)$ GeV and $m_{\chi_1^0}=100$ GeV.  

\item{$\tilde{t}_2 \rightarrow \tilde{t}_1 Z/h$~\cite{Aad:2014mha,Khachatryan:2014doa, CMS-PAS-SUS-13-021}}

Searches for the second stop utilize the decay of $\tilde{t}_2 \rightarrow \tilde{t}_1 Z/h$, looking for signals including $b$-jets and large $\met$ with either same flavor leptons reconstruction of the $Z$ boson \cite{Aad:2014mha} and/or high $p_T$ jet and $b$-jet multiplicities with additional leptons  \cite{Khachatryan:2014doa, CMS-PAS-SUS-13-021}. The interpretation is performed in the region $m_{\tilde{t}_1}-m_{\chi_1^0}\sim m_t$, which is hard to probe by $\tilde{t}_1\rightarrow t \chi_1^0$ channel given the relative small $\met$.   For ${\rm BR}(\tilde{t}_2\rightarrow \tilde{t}_1 Z)=100\%$, the second stop mass is excluded up to about 600 GeV for a light LSP mass.   For ${\rm BR}(\tilde{t}_2\rightarrow \tilde{t}_1 h)=100\%$, the second stop mass exclusion limit is about 540 GeV.   When the decay branching fraction to $\tilde{t}_1Z$ and $\tilde{t}_1h$ is 50\% each, the exclusion limit is about 580 GeV for a light LSP mass.

\item{$\tilde{b}_1\rightarrow b \chi_1^0$~\cite{Aad:2014nra,Aad:2013ija,CMS-PAS-SUS-13-018}}

Sbottom pair production with $\tilde{b}_1\rightarrow b \chi_1^0$ leads to signals with two $b$ jets and large $\met$.  The null results from   ATLAS~\cite{Aad:2013ija} exclude sbottom masses up to 620 GeV for $m_{\chi_1^0}<$120 GeV.     $m_{\tilde{b}_1}-m_{\chi_1^0}$ is excluded up to 50 GeV for  sbottom masses  up to 300 GeV.    The exclusion limits depend sensitively on the branching fraction of $\tilde{b}_1 \rightarrow b \chi_1^0$.  For 60\% branching fractions, the sbottom exclusion limit is reduced to 520 GeV.   The CMS exclusion limits are about 70 GeV stronger~\cite{CMS-PAS-SUS-13-018}.   For small mass splitting between sbottom and the LSP: $m_{\tilde{b}_1}-m_{\chi_1^0}\sim m_b$, monojet plus $\met$ search excludes  sbottom masses up to about 255 GeV~\cite{Aad:2014nra}.

\item{$\tilde{b}_1\rightarrow b \chi_2^0$~\cite{Aad:2014lra,CMS-PAS-SUS-13-008}}
  
Sbottom searches on direct sbottom pair production with  $\tilde{b}_1\rightarrow b \chi_2^0$ with 100\% decay branching fraction of $\chi_2^0\rightarrow \chi_1^0 h$ have been performed at ATLAS~\cite{Aad:2014lra}, searching for signals with zero lepton, large $\met$,  high jet multiplicity and at least three $b$-tagged jets.   Null search results exclude the sbottom masses between 340 and 600 GeV for $m_{\chi_2^0}=300$ GeV and $m_{\chi_1^0}=60$ GeV.   No sensitivity is obtained for $m_{\chi_2^0}<240$ GeV due to the soft $\met$ in the signal events.    For $\tilde{b}_1\rightarrow b \chi_2^0$ with 100\% decay branching fraction of $\chi_2^0\rightarrow \chi_1^0 Z$, three leptons plus one $b$ jet plus $\met$ search at the CMS excludes sbottom masses up to 450 GeV for   LSP masses  between 100 to 125 GeV and $m_{\chi_1^\pm}-m_{\chi_1^0}=110$ GeV~\cite{CMS-PAS-SUS-13-008}.

 \item{$\tilde{b}_1\rightarrow t \chi_1^\pm$~\cite{CMS-PAS-SUS-13-008,Chatrchyan:2013fea,Aad:2014pda,Chatrchyan:2014aea}}
  
 Sbottom searches on direct sbottom pair production with  $\tilde{b}_1\rightarrow t \chi_1^\pm$ with 100\% decay branching fraction of $\chi_1^\pm\rightarrow W \chi_1^0 $ have been performed at both ATLAS and CMS~\cite{Aad:2014pda,Chatrchyan:2013fea}, looking  for signals with two same charge leptons or three leptons plus multiple jets.  The interpretation was done for fixed $m_{\chi_1^0}=60$ GeV as well as varying $m_{\chi_1^0}$ with $m_{\chi_1^\pm}=2m_{\chi_1^0}$.   The sbottom mass limit is about 440 GeV in both cases for $m_{\chi_1^\pm} < m_{\tilde{b}_1}-m_t$~\cite{Aad:2014pda}.  The CMS limits are about 50 to 100 GeV stronger~\cite{Chatrchyan:2013fea,CMS-PAS-SUS-13-008,Chatrchyan:2014aea}.

 \end{itemize}

At the 14 TeV LHC, with the dominant decay channel of $\tilde{t}_1\rightarrow t \chi_1^0$, studies using semileptonic channel and fully hadronic channel show that  for   LSP masses below 200 GeV, a 5$\sigma$ reach of stop discovery is possible for stop masses  up to about 1 TeV with 300 ${\rm fb}^{-1}$ integrated luminosity~\cite{ATL-PHYS-PUB-2013-011}.  For the high luminosity option of LHC (HL-LHC) with  3000 ${\rm fb}^{-1}$ integrated luminosity, the discovery reach is extended by about 200 GeV.   The 95\% exclusion limit is about 1.2 TeV (1.45 TeV) with 300 (3000) ${\rm fb}^{-1}$ integrated luminosity.  For sbottom searches with $\tilde{b}_1\rightarrow b \chi_1^0$, the discovery reach is   about 1.1 (1.3) TeV and the exclusion reach is about 1.4 (1.55) TeV with 300 (3000) ${\rm fb}^{-1}$ integrated luminosity~\cite{ATL-PHYS-PUB-2014-010}.  CMS analyses using specific full spectrum benchmark points show similar sensitivities~\cite{CMS-PAS-SUS-14-012}.

\section{Collider analysis}
\label{sec:analyses}

 Given a different neutralino/chargino mass spectrum, many new decay channels for stop and sbottom appear, while the channels of $\tilde{t}_1\rightarrow t \chi_1^0, b\chi_1^\pm$ and $\tilde{b}_1\rightarrow b \chi_1^0$ could be highly suppressed.   This   leads to the relaxation of current collider search limits based on those above mentioned channels.  At the same time, those new channels provide new discovery  opportunities.   To demonstrate the new discovery potential, we pick one particular channel as our benchmark scenario for collider analyses.  Studies on other possible mass spectrum and decay channels are left for future study. 

In this section, we study the detectability of   the light stop   in Case IA with a mass hierarchy of $M_1 < M_2 < M_{3SQ} \ll |\mu|, M_{3SU}$,  resulting in a mass spectrum including a mostly left-handed stop and mostly left-handed sbottom, Wino-like NLSPs, and a Bino-like LSP.     In our analyses, we consider the kinematic region of $m_{\tilde{t}_1}-m_{\chi_2^0}>m_t$ and $m_{\chi_2^0}-m_{\chi_1^0}>m_h$ such that  $\tilde{t}_1 \to t {\chi}_2^0$ and 
 ${\chi}_2^0 \to h {\chi}_1^0 $ are kinematically open.     The collider analyses of the current event topology can not be applied for the more compressed scenarios when either $M_{3SQ}$ is close to $M_2$ or $M_2$ is close to   $M_1$.   To illustrate the decay branching fractions, we choose a benchmark point with the specific set of parameters and  the corresponding mass spectrum shown in Table~\ref{table:MassParameters}.  The value of $\tilde{A}_t$    is chosen such that the SM-like Higgs mass is around 125 GeV.   Note that even though $\tilde{A}_t$ is set to a large value, the large mass splitting between $M_{3SQ}$ and $M_{3SU}$ results in  a mostly left-handed ${\tilde t}_1$ and  mostly right-handed ${\tilde t}_2$.  Therefore,   the decay patterns of $\tilde{t}_1$ and $\tilde{b}_1$ follow those of the Case IA: purely left-handed stop/sbottom in the minimal mixing scenario.

\begin{table}[ht]
\begin{tabular}{|c|c|c|c|c|c||c|c|c|c|c|c|} \hline
    $M_1$ & $M_2$   & $\mu$ & $\tilde{A}_t$ & $M_{3SQ}$ & $M_{3SU}$ &${\chi}_1^0$ & ${\chi}_2^0$ & ${\chi}_1^+$ & $\tilde{t}_1$ & $h$ & $\tilde{b}_1$\\ \hline
     150  & 300   &  2000 & 2750 &  650  & 2000 &  151 & 319 & 319 & 646 & 125 & 637 \\ \hline
\end{tabular}

\caption{Mass parameters and mass spectrum of SUSY particles for the benchmark point.  All masses are in units of GeV.   
 }
\label{table:MassParameters}
\end{table}

 The decay channels for the light stop of the benchmark point are shown in Table~\ref{table:decaychannel}.  While the dominant decay channel is  $\tilde{t}_1 \to b {\chi}_1^+$ with 71\% branching fraction, the subdominant channel $\tilde{t}_1 \to t {\chi}_2^0$ is  about 27\%,    providing an interesting signal where   ${\chi}_2^0$ can either decay to a Higgs   or a $Z$ boson.      For our choice of parameters with $\mu>0$, ${\chi}_2^0$ dominantly decays to $h {\chi}_1^0$, as shown in Table~\ref{table:decaychannel}.  Flipping the sign of $\mu$ could lead to another interesting channel of $\chi_2^0 \rightarrow Z \chi_1^0$, which is left for future study.

\begin{table}[ht]
\centering
       \begin{tabular}{|c|c|} \hline
    Decay & Branching Fraction  \\ \hline
    $\tilde{t}_1 \to t {\chi}_1^0$  & 2\%   \\ \hline
    $\tilde{t}_1 \to t {\chi}_2^0$  & 27\%   \\ \hline
    $\tilde{t}_1 \to b {\chi}_1^+$  & 71\%   \\ \hline
   \end{tabular}
   \begin{tabular}{|c|c|} \hline
    Decay & Branching Fraction  \\ \hline
    ${\chi}_2^0 \to Z {\chi}_1^0 $ &  3\% \\ \hline
    ${\chi}_2^0 \to h {\chi}_1^0 $ & 97\% \\ \hline
    ${\chi}_1^+ \to W^+{\chi}_1^0  $ & 100\% \\ \hline

   \end{tabular}

\caption{Decay branching fractions  of $\tilde{t}_1$, ${\chi}_2^0$ and ${\chi}_1^+$ for the benchmark point.  }
\label{table:decaychannel}
\end{table}

 For our benchmark point with the reduced branching fraction of  ${\rm BR}(\tilde{t}_1 \to b \chi_1^\pm)=$ 71\%, the current collider search limits on the stop are much more relaxed: less than about  500 GeV for $m_{\tilde{t}_1}$.    However, new search channels open up, which play a complementary role for stop searches at the LHC. 

In our analysis, we study the  stop pair production with  mixed stop decay final states of  $\tilde{t}_1 \to t {\chi}_2^0  \to th {\chi}_1^0$,  $\tilde{t}_1 \to b {\chi}_1^\pm  \to bW {\chi}_1^0 $.  The branching fraction for such decay is about 38\% for our benchmark point and varies between 25\% and 50\% for a stop mass larger than 500 GeV with $M_2=M_1+150$ GeV.      We consider semileptonic decays of  the two $W$s  and the Higgs decay to two $b$-quarks.   Since we choose the CP-odd Higgs mass $m_A$ to be 2000 GeV, we are in the decoupling region of the Higgs sector with the light CP-even Higgs being SM-like.  Given that we are in the Bino-LSP scenario with $M_2=M_1+150$ GeV, additional possible decay modes of $h$ into neutralino/charginos are either highly suppressed or kinematically forbidden.   Therefore, the light CP-even Higgs is consistent with   the observed  signal of a 125 GeV SM-like Higgs boson.  In our analyses, we have taken the branching fraction of $h\rightarrow b\bar{b}$ to be  the SM value of $57.7\%$.    The signal contains   four $b$-jets, two jets, one isolated lepton and large missing energy. The presence of a single lepton helps to reduce QCD multijets backgrounds without significant branching fraction suppression.

  The dominant SM backgrounds are $bbWW$ (dominantly from $t\bar{t}$) and $t\bar{t}b\bar{b}$.   While $t\bar{t} h$ is an irreducible background, the production cross section is typically small.   Other backgrounds consist of $t\bar{t} W$ and $t\bar{t} Z$.  

Event samples are generated  using Madgraph MG5$\_$aMC$\_$V2$\_$2$\_$1 \cite{Alwall:2014hca},   processed   through Pythia 6.420 \cite{oai:arXiv.org:hep-ph/0603175} for   fragmentation and hadronization and then through Delphes-3.1.2 \cite{deFavereau:2013fsa} with the Snowmass combined LHC detector card  \cite{Anderson:2013kxz} for  detector simulation.  Both the SM backgrounds and the stop pair production signal are normalized to theoretical cross sections, calculated including higher-order QCD corrections \cite{oai:arXiv.org:1006.4771, Broggio:2013uba, Borschensky:2014cia,oai:arXiv.org:0804.2800, oai:arXiv.org:0905.0110, oai:arXiv.org:hep-ph/0211352, oai:arXiv.org:1204.5678, oai:arXiv.org:0804.2220}.     For event generation, we have set $m_t$ to be 173 GeV, and the Higgs mass $m_h$ to be 125 GeV.   The renormalization scale and factorization scale are taken to be $\sqrt{M^2 + p_T^2}$ for a single heavy particle.  For pair production of  heavy particles,  the geometric mean of $\sqrt{M^2 + p_T^2}$ for each particle is used.   For the signal process, we scan the parameter range of $M_{3SQ}=400 \ldots 1100$ GeV with step size of 25 GeV, and $M_1= 3 \ldots 750$ GeV with step size of  25 GeV.  We fix $M_2$ to be  $M_2=M_1+$ 150 GeV.

 We apply the following basic event selection cuts:

\begin{itemize}
\item Events are required to have at least four isolated jets\footnote{ The anti-$k_t$ jet algorithm is used in the reconstruction of jets, with the jet radius being 0.5.  For isolated jets, we require any jet within $\Delta R$ = 0.2 of a lepton   be discarded.   
An event is discarded if the distance between $\met$ and all jets, $\Delta \Phi(\met,j)  $, is less than  0.8.  } with
\begin{equation}
p_T^{j1, j2, j3} > 40 \ {\rm GeV}, \ p_T^{j4 } > 25 \ {\rm GeV}, \ |\eta^j| < 2.5.
\end{equation}
All isolated jets   satisfying $p_T^{j} > 25 \ {\rm GeV}, \ |\eta^j| < 2.5$ are   counted in $N_j$.
 
\item  Among the jets, at least two are $b$-tagged jets.    The $b$-tagging efficiency depends on the $p_T$ and $\eta$ of the jets,  which is 0 for $p_T<15$ GeV or $|\eta|>2.5$,  about 70\% for  $|\eta^j| < 1.2$ and about 60\% for $1.2<|\eta^j| < 2.5$ with $p_T^j\gtrsim 200$ GeV.  The mistag rate depends on the quark species, as well as $p_T$ and $\eta$ of the jets.  It is about 15\% for $c$-quark and a constant 2\% for light jets.

\item One isolated lepton\footnote{ For an isolated lepton, we require $\Delta R(\ell,j) > 0.4$.} ($\textit{e}$ or $\mu$) is required to have
\begin{equation}
p_T^\ell > 20 {\rm \ GeV \ with} \ |\eta^\ell| < 2.5.
\end{equation}

 \end{itemize}

Additional optimization selection cuts are applied to  further  enhance the signal and suppress the SM backgrounds:

 \begin{itemize}
\item $\met$, defined as the magnitude of the missing transpose momentum, $\mathbf{p}^{miss}_T$,   to be  above 100, 120, 140, 160 180, and 200 GeV.   
\item $H_T$,   defined as the scalar sum of the $p_T$ of all surviving isolated jets   satisfying  $p_T^{j} > 25 \ {\rm GeV}, \ |\eta^j| < 2.5$: $H_T = \sum p_T^{jet}$,   to be  above  400, 450, 500, 550, 600 GeV. 
\item Transverse mass $m_T$, defined as the invariant mass of the lepton and the missing transpose momentum: 
\begin{equation}
m_T = \sqrt{2  p_T^\ell\met (1-\cos\phi(\mathbf{p}_T^\ell,\mathbf{p}^{miss}_T))}, 
\end{equation}
 to be  above 100, 120, 140, 160, 180, 200 GeV.
\item $N_j$,   the number of all surviving isolated jets  satisfying  $p_T^{j} > 25$ GeV  and  $|\eta^j| < 2.5$,  to be at least 4, 5, or 6.
\item $N_{bj}$,   the number of all tagged $b$-jets, to be at least 2, 3, or 4. 
\end{itemize}

\begin{figure}[t]
\begin{center}
\includegraphics[width = 0.48 \textwidth]{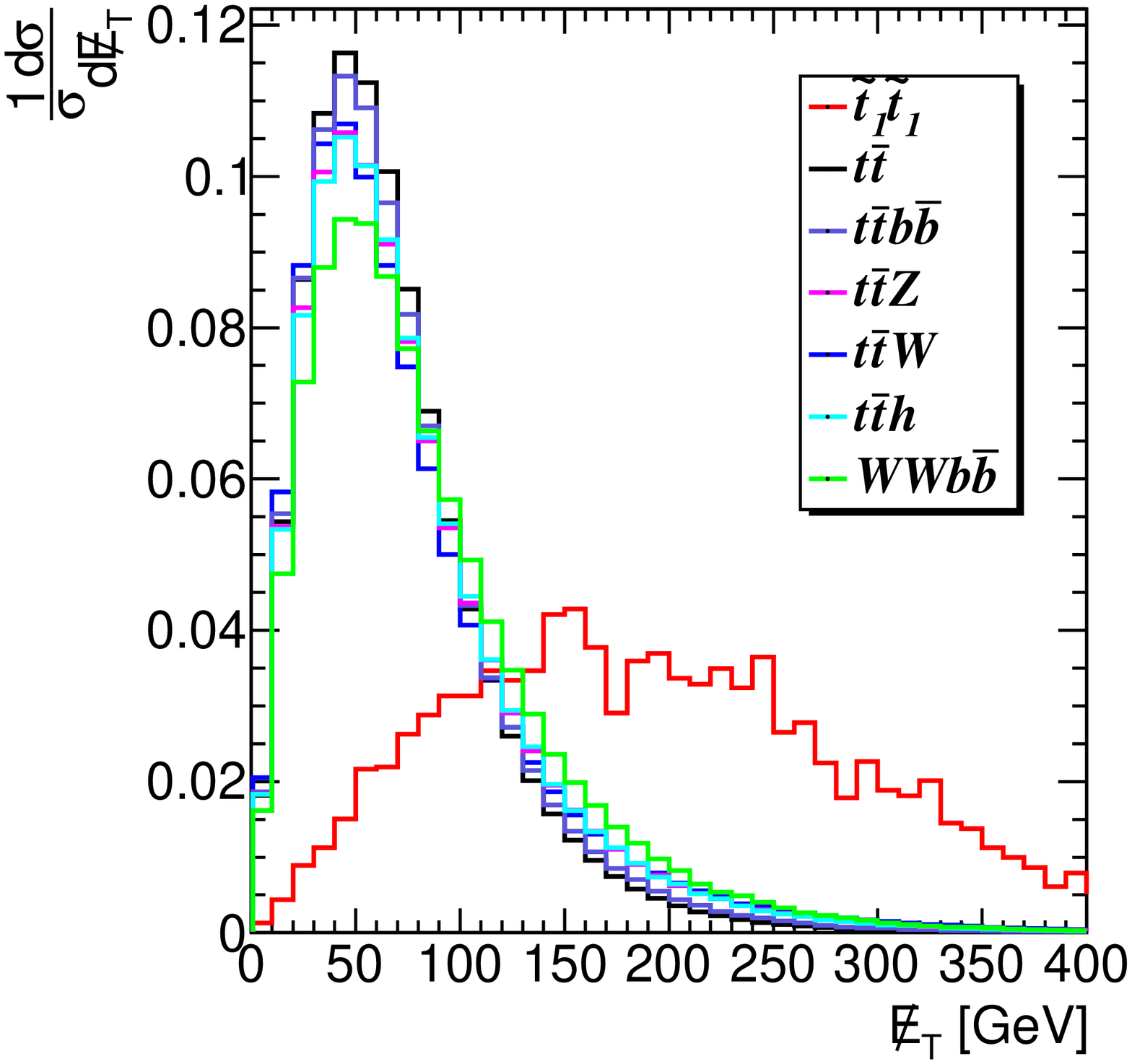}
\includegraphics[width = 0.48 \textwidth]{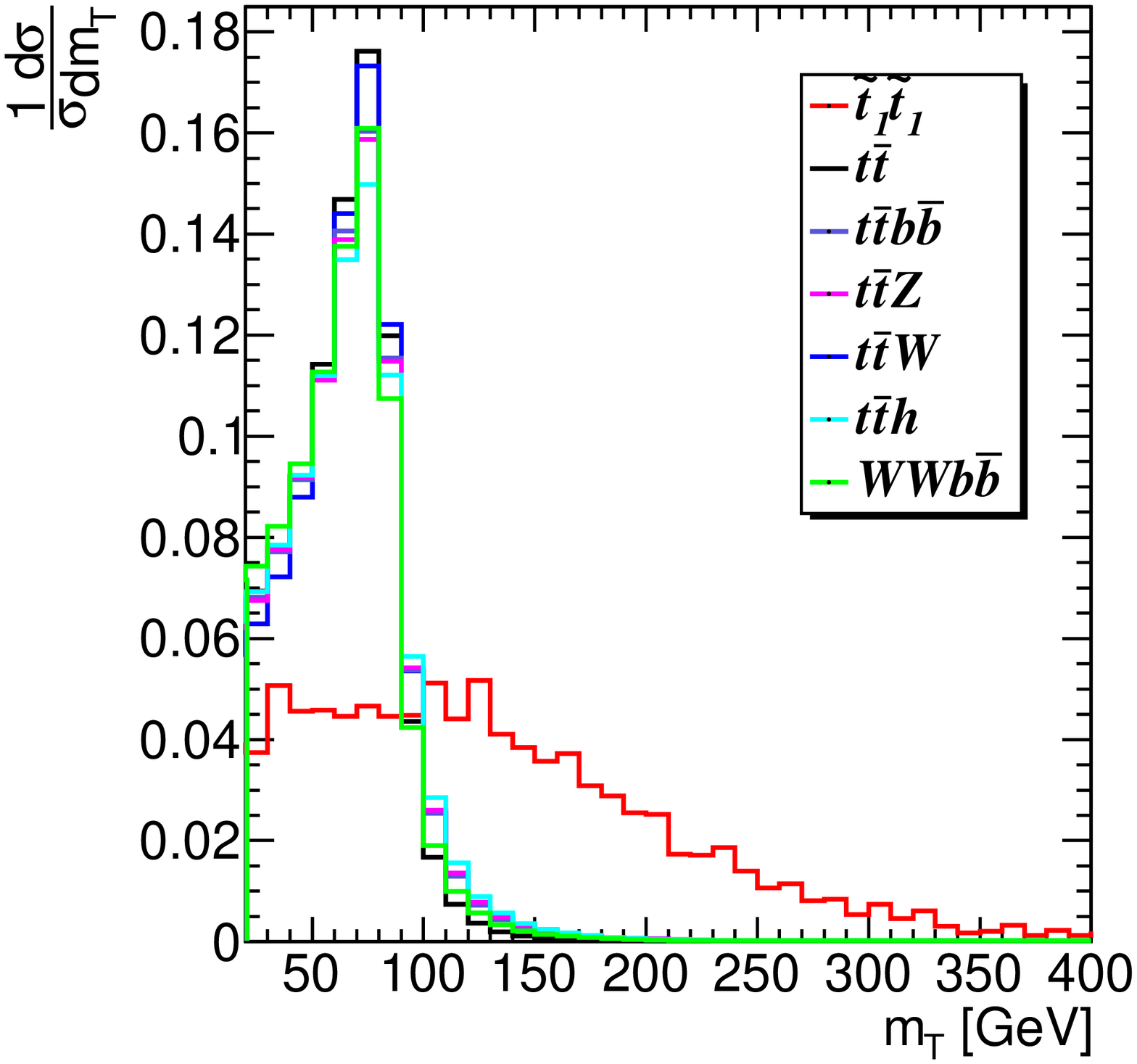}
  \end{center}
\caption{The  distribution of $\met$ (left)  and $m_T$ (right)  for the signal at the benchmark point and the SM backgrounds  after basic selection cuts.  }
\label{Figure:distribution}
\end{figure}

The distributions of $\met$ and $m_T$ for both the signal and   the SM backgrounds are shown in Fig.~\ref{Figure:distribution}. In the $\met$ distribution,  the $\met$ for all the SM backgrounds comes only from the neutrino, which is typically smaller than that of the signal with additional $\met$ contribution from the LSP.   The transverse mass for the signal process extends beyond the SM threshold of the $W$ boson mass.  The $H_T$ distribution of the signal is maximum at a higher value compared to the SM backgrounds.

In Table~\ref{table:cutefficiency}, we present the cumulative cut efficiencies for the signal and dominant SM backgrounds with one set of selection cuts. By utilizing  strong $\met$, $H_T$ and  $m_T$ selection cuts, we significantly reduce the SM backgrounds.    The stop signal process typically generates multiple hard jets in our specified decay.  The $N_{bj}$ cut further plays an important role in cutting $t\bar t$, $t\bar{t}W$, and $t\bar{t}Z$  backgrounds.  $t \bar t$ is the dominant background given  its large cross section.   The tails in the $t\bar{t}$ missing $E_T$ and $H_T$ distribution are more relevant than those of the rare SM processes of $ttZ/W$.  $t\bar{t}b\bar{b}$ is the  second dominant background given its relatively large cross section and similar final states to the signal process.  $t\bar{t}h$, $t\bar{t}Z$, and $t\bar{t}W$   can be sufficiently suppressed due to low cross sections.   We impose a constraint on the number of signal events, $N_s \ge$ 3 for 300 ${\rm fb}^{-1}$ in order to obtain sufficient statistics.   

\begin{table}[ht]
\centering
%\subfloat[list1][Decay Channels]{
  %\rule{4cm}{3cm}
     \begin{tabular}{|c|c|c|c|c|c|c|} \hline
    Description & $\tilde{t}_1 \tilde{t}_1$   & $t\bar t$ & $t\bar t b \bar b$   & $t \bar t h$ & $ t \bar tZ$ &$t\bar{t}W$ \\ \hline
    CS (fb) & 10  & 261230 & 2346   & 108 & 221 & 218 \\ \hline
    %Original  & 80,000,000 & 2,500,000 & 1,250,000 & 1,250,000 & 1,250,000 &1,500,000 & 10,000 \\ \hline
    Basic selections & 39\% & 14\% & 24\% & 31\% & 30\% & 25\%\\ \hline
    %Basic cuts  & 12,343,398 & 946,129 & 489,301 & 351,892 & 168,477 & 519,388 & 4820 & 4426 \\ \hline
    %$E_T^{miss} >$ 200 & 207,065 & 29274 & 20,182 & 17,898 & 8,742 & 22,543 & 1479 & 1063 \\ \hline
    $\met >$ 200 GeV & 18.5\%  & 0.23\% & 0.58\%  & 1.2\%& 1.2\%& 1.2\%  \\ \hline
    %$H_T >$ 500 & 67,463 & 17,611 & 12,779 & 9,997 & 3,245 & 12,849 & 952 & 641 \\ \hline
    $H_T >$ 500 GeV & 15.6\%  & $7.4\times10^{-4}$ & 0.29\% & 0.78\% & 0.77\%& 0.69\% \\ \hline
    %$m_T >$ 150 & 211 & 362 & 93 & 64 & 34 & 79 & 360 & 231 \\ \hline
    $m_T >$ 160 GeV & 5.9\%  & $1.8\times10^{-6}$ & $3.6\times10^{-5}$ & $6.6\times10^{-5}$ & $7.0\times10^{-5}$& $6.0\times10^{-5}$ \\ \hline
    %$N_j \ge$ 5 & 134 & 227 & 48 & 19 & 9 & 35 & 271 & 185 \\ \hline
    $N_j \ge$ 5 & 4.4\%  & $8.5\times10^{-7}$ & $2.1\times10^{-5}$ & $3.7\times10^{-5}$& $3.8\times10^{-5}$& $2.6\times10^{-5}$ \\ \hline
    %$N_{bj} \ge$ 2 & 27 & 118 & 24 & 4 & 1 & 10 & 146 & 101 \\ \hline
    $N_{bj} \ge$ 2 & 2.9\% & $2.9\times10^{-7}$ & $1.1\times10^{-5}$ & $2.2\times10^{-5}$& $1.1\times10^{-5}$  & $7.6\times10^{-6}$ \\ \hline
    CS (fb) after selection cuts & 0.29  & 0.075 & 0.026  & 0.0023 & 0.0025 & 0.0017 \\ \hline
  \end{tabular}

%}
\caption{The cumulative cut efficiencies for the signal at the benchmark point and all SM backgrounds.  The cross sections shown in the second row are for the semileptonic final states.  }
\label{table:cutefficiency}
\end{table}

\begin{figure}[t]
\begin{center}
\includegraphics[width = 0.60 \textwidth]{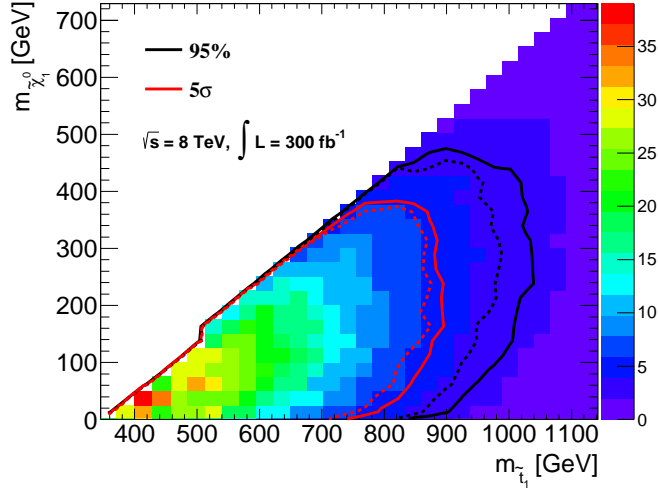}
  \end{center}
\caption{The plot shows the 5$\sigma$ discovery reach (red) and 95\% exclusion limits (black) of the stop  in the $m_{\tilde{t}_1 }-m_{\n}$ plane for  14 TeV  LHC with 300 ${\rm fb}^{-1}$ of integrated luminosity.     $M_2$ is fixed to be  $M_1 + 150$ GeV and  10\% (30\%) systematic uncertainties are assumed for solid (dotted) curves.   The color coding on the right indicates the signal significance defined simply as $S/\sqrt{B}$ to guide the eye.  For exclusion and discovery reach, we   used the signal significance defined as $S/\sqrt{S+B+(\epsilon\times B)^2}$ and $S/\sqrt{B+(\epsilon\times B)^2}$, respectively.  Here $\epsilon$ is the assumed systematic uncertainty.}   
\label{Figure:reach_plot}
\end{figure}

In  Fig.~\ref{Figure:reach_plot}, we show the 95\% C.L. exclusion limit and 5$\sigma$ reach  in the parameter space of $m_{\tilde{t}_1}$  versus $m_{\chi_1^0}$ for the 14 TeV LHC with 300 ${\rm fb}^{-1}$ luminosity.    $M_2$ is fixed to be  $M_1 + 150$ GeV and   10\% (30\%)  systematic uncertainties  on SM backgrounds are assumed for solid (dotted) curves.     For each mass point of ($m_{\tilde{t}_1}, m_{\chi_1^0})$, given the mass dependence of the production cross section and decay branching fractions, the signal $\sigma\times{\rm BR}$ for each individual point has been used.  All combinations of the cut values for the advanced selection cuts of $\met$, $H_T$, $m_T$, $N_j$ and $N_{bj}$ are examined.  The optimized combination that gives the best significance  is used for that particular mass point.    For the 5$\sigma$ reach, stop masses up to 740 GeV can be reached for a massless LSP and about  940 GeV  with $m_{{\chi}_1^0}$ = 250 GeV,  assuming 10\% systematic uncertainties. The 95\% C.L. exclusion limits  are about 840 GeV for stops with a light $\chi_1^0$, while the reach is  1040 GeV for $m_{{\chi}_1^0} =$ 250 GeV.  Limits with 30\% systematic uncertainties  are about 50 GeV worse.

 Note that a light left-handed sbottom with mixed decay of $\tilde{b}_1\rightarrow b \chi_2^0$ and $\tilde{b}_1\rightarrow t \chi_1^\pm$ could lead to the same final states.  We focused on the stop search sensitivities in the current study.  Collider studies for the sbottom search as well as the combined reach in $M_{3SQ}$ versus $m_{\chi_1^0}$ plane can be found in Ref.~\cite{sbottom_future}.

\section{Summary and Conclusion}
\label{sec:conclusion}

Most of the current stop and sbottom searches at the LHC have been performed considering the channels of $tt+\met$, $bbWW+\met$ for stop and $bb+\met$ for sbottom,  assuming the stop and sbottom decay 100\% into these channels.     However, in many regions of MSSM parameter space, these decay channels are subdominant, resulting in relaxed bounds from current LHC searches.    In this work, we studied decays of the stop and sbottom in the cases of a Bino-like LSP with either   Wino-like or Higgsino-like NLSPs in the low energy spectrum, for the left- and right-handed stops and left-handed sbottom in the minimal mixing scenario, and $\tilde{t}_{1,2}$, $\tilde{b}_1$ in the maximal mixing scenario.  We found that  new decay channels of $\tilde{t}_1\rightarrow t \chi_{2,3}^0$,  $\tilde{b}_1\rightarrow b \chi_{2,3}^0, t\chi_1^\pm, W\tilde{t}_1$ open up,  which could even dominate over $\tilde{t}_1\rightarrow t\chi_1^0, b\chi_1^\pm$ and $\tilde{b}_1\rightarrow b \chi_1^0$ channels.   For the heavier stop state,  $\tilde{t}_2$,  a new channel of $\tilde{t}_2 \rightarrow W \tilde{b}_1$ appears in addition to $\tilde{t}_2 \rightarrow Z \tilde{t}_1$ in the maximal mixing scenario.   Given the further decays of $\chi_{2,3}^0$ and $\chi_1^\pm$, pair production of stops and sbottoms at the LHC typically leads to $bb$ plus multiple gauge bosons plus $\met$ final states.   Current   search channels of $bbWW+\met$ and $bb+\met$ could be highly suppressed.

We performed a sample collider analysis for the reach of the stop  at the 14 TeV LHC with 300 ${\rm fb}^{-1}$ integrated luminosity for one particularly interesting channel in the  Bino-like LSP with Wino-like NLSP case. We considered left-handed stop pair production  mixed stop decay final states of  $\tilde{t}_1 \to t {\chi}_2^0  \to th {\chi}_1^0$,  $\tilde{t}_1 \to b {\chi}_1^\pm  \to bW {\chi}_1^0 $, leading to the $bbbbjj\ell+\met$ collider signature.  The branching fraction for such decay varies between 25\% and 50\% for a stop mass larger than 500 GeV with $M_2=M_1+150$ GeV.    Our results show that   for a LSP mass of 250 GeV, the 95\% C.L. exclusion reach is about 1040 GeV for the stop and the 5$\sigma$ reach is about 940 GeV,  assuming 10\% systematic uncertainties.     The reach decreases with smaller LSP mass. 
 
Considering different low-lying neutralino/chargino spectra provides several promising channels for the stop and sbottom study.  In this paper we focused on final states with a Higgs boson.  Decays of $\chi_2^0$ to $Z \chi_1^0 $  could   be dominant with a different choice of sign($\mu$).    Furthermore, a different mass spectrum of neutralino/chargino with LSP being either Wino-like or Higgsino-like might give rise to more interesting final states. It  is important to identify the leading decay channels in various regions of parameter space to fully explore the reach of the LHC for the third generation squarks, which has important implications for the stabilization of the electroweak scale  in supersymmetric models.  The strategy developed in our analysis can   be applied to the study of  top partners in other new physics scenarios as well.

\acknowledgments
We would like to thank Felix Kling, Tao Han, Yongcheng Wu, Bing Zhang  for helpful discussions.   The work is supported by the  Department of Energy  under Grant~DE-FG02-04ER-41298.  This work was supported in part by National Science Foundation Grant No. PHYS-1066293 and the hospitality of the Aspen Center for Physics.

 \bibliography{bibliography}
 
\end{document}